\documentclass[journal=cmatex,manuscript=article]{achemso}
\usepackage{achemso}

\setkeys{acs}{maxauthors = 0}
\usepackage{setspace}
\usepackage{amssymb,amsmath}
\usepackage{graphicx}
\usepackage{float}
\usepackage{multirow}
\usepackage{booktabs}
\usepackage{chemformula}
\usepackage[version=3]{mhchem}
\usepackage{xcolor}
\usepackage{xcolor}

\usepackage{hyperref} 
\hypersetup{
     colorlinks   = true,
     citecolor    = blue,
     linkcolor   = blue
}
\graphicspath{{./Images/}}
\usepackage{siunitx}

\title{Best Practices for First-Principles Modeling of Amorphous Oxide Semiconductors: A Statistical Framework and Application to Zn-Sn-O}

\author{Michiel J. van Setten} % 0000-0003-0557-5260
\email{michiel.vansetten@imec.be} 
\affiliation{imec, Kapeldreef 75, Leuven, Belgium}
\altaffiliation{ETSF, European Theoretical Spectroscopy Facility}
\author{Tonglin L. Newsom}
\affiliation{Electrical Engineering and Computer Science Department, University of Michigan, Ann Arbor, MI, USA}
\author{Christopher Pashartis}
\affiliation{imec, Kapeldreef 75, Leuven, Belgium}
\author{Vera van Noort}
\affiliation{KU Leuven, Department of Microbial and Molecular Systems, Leuven, Belgium}
\altaffiliation{Leiden University, Institute of Biology Leiden, The Netherlands}
\author{Rebecca L. Peterson}
\affiliation{Electrical Engineering and Computer Science Department, University of Michigan, Ann Arbor, MI, USA}
\author{Geoffrey Pourtois} % 0000-0003-2597-8534
\affiliation{imec, Kapeldreef 75, Leuven, Belgium}

\keywords{semiconductor, doping, first principles, amorphous phase}

\begin{document}

\begin{abstract}
Ternary and quaternary amorphous oxide semiconductors have many properties that make them promising candidates for use in electronic applications like display, memory, and back end of line logic. However, finding the right material for a given application and optimizing its properties, deposition, and integration, requires a thorough understanding of the physics and chemistry at play. When properly carried out, first principles computations can play a crucial role in enhancing this understanding. In this work, we highlight several pitfalls often observed in research applying these computations, with the Zn-Sn-O system as an example. We show that a proper understanding of the fundamental differences between the physics of the crystalline and amorphous or disordered phases is crucial, as is a proper statistical sampling of structural models. For the Zn-Sn-O system we conclude that from a performance point of view, mobility and initial threshold voltage, it is a promising material class. However, our computed results show that a similar sensitivity to hydrogen induced doping may be present as in IGZO.
\end{abstract}

\maketitle

\section{Introduction}

The core of each advanced integrated circuit consists of a layer of the smallest transistors imaginable, directly fabricated with sophisticated deposition and lithography steps from a layer of pure silicon substrate. The heart of these transistors, the channel, is the current-carrying, active layer that makes the transistor switch. It is made from the substrate itself or, more recently, from an epitaxially grown layer of silicon-germanium on top of the silicon substrate. In many decades of research, these materials have never been beaten at this job. However, if we want to keep decreasing the transistor footprint and energy consumption, we need to move to more complex architectures. Many of these, like back end of line logic, but also applications like flexible electronics, do not allow for the temperatures needed to create that very pure, crystalline layer.  The further advancement of semiconductor devices, needs new materials with more moderate deposition conditions. One possibility is to move from the crystalline to the amorphous phase.\cite{Nomura2004, Walsh2009, Banger2010, Park2012, Zeng2024} Unfortunately, the amorphous phase of silicon, which can be deposited at milder conditions, has inferior electrical properties.\cite{Powell1989}

As an alternative to amorphous silicon, many amorphous oxide semiconductors have been investigated.\cite{Buchholz2014, Sil2022} The hybridization of metal-s, oxygen-s, and oxygen-p atomic orbitals forming the bottom of the conduction band gives these materials a decent electron mobility even in the amorphous phase,\cite{Walsh2009} surpassing that of amorphous silicon.\cite{Medvedeva2020} One of the most famous oxide semiconductors used in the amorphous phase is \ch{InGaZnO4} (IGZO), which can work as the active layer in n-type thin film transistors featuring extreme low off-currents. These features allowed for its use in display and memory applications.\cite{Kamiya2010, Arai2011, Kataoka2013, Nag2013, Chasin2014, Nadarajah2015, Nag2017, DeRoose2017, Mo2019, Sharifi2020, Belmonte2020, Hiblot2021, Han2021, Liu2024}

Amorphous IGZO, unfortunately, is also not ideal. All three involved metals, in particular In, are on the list of endangered elements as published by the American Chemical Society.\cite{endangered_elements} The electron mobility is on the low side and the material is chemically not very stable. For instance, the interaction with various metals can cause oxygen scavenging, leading to additional doping or defects.\cite{Mitard2020, vanSetten2021_odef_igzo, Kruv2025} Moreover, the interaction with hydrogen, during deposition, integration, or operation, can also lead to uncontrolled doping.\cite{Kim2014, deJamblinnedeMeux2015, deJamblinnedeMeux2017, deJamblinnedeMeux2018, Medvedeva2020, Kljucar2020, Vogt2020, vanSetten2021_odef_igzo, Kang2021, Subhechha2022, Wu2022, Rinaudo2023} In a transistor, this excess doping of the active layer leads to an unwanted shifting of the switching voltage. In order to resolve this, many studies have been dedicated to finding better alternatives, and even in last few years new materials have been proposed.\cite{Choi2020, Takahashi2020, Cha2020, Lestari2020, Guo2020, Jeong2020, AvelarMuoz2020, Park2020} Systematic screenings and general guidelines and insight development, however, are still limited.\cite{Janotti2009, Krner2015, Moffitt2017, Thyagarajan2020, Zhang2020, Shiah2021, vanSetten2022screening, Xu2023, Kruv2023, Choi2024}

An interesting candidate is the Zn-Sn-O system. It does not contain indium, which poses a cost and availability advantage. Moreover, it combines a low deposition temperature, as low as 180$^\circ$C, \cite{Fernandes2018, Allemang2020} with promising electronic properties. In the fully oxidized, Zn$^{2+}$ Sn$^{4+}$, stoichiometric state we have for each Zn one and for each Sn two oxygen atoms. In this state, as we confirm in this work, the conduction states become a hybridization of metal-s, oxygen-s, and oxygen-p, which like in IGZO are strongly delocalized, which is a necessary condition for good mobility. The valence states are, as in IGZO, built up from Zn-d and O-p states, causing strong charge carrier localization. The latter, combined with the large bandgap, is helpful to create a low off-current in n-type transistors since minority transport is suppressed. An important additional feature that may cause the Zn-Sn-O system to outperform IGZO in terms of performance is the stability of its electronic gap with respect to compositional variation, as we will illustrate in this paper.

% DFT literature
Several first principles computational studies using density functional theory (DFT) on the Zn-Sn-O system are present in literature. Unfortunately, many of the mistakes and shortcuts, commonly made in computational studies of amorphous oxides, also cloud the understanding of this material. One of the main issues is the lack of statistical sampling of the morphology and coordination space that comes with the amorphous phase. A periodically repeated unit cell is usually used to mimic a true macroscopic amorphous material. Due to its size this is often called a super cell. The largest, reasonably-computable super cell contains in the order of one thousand atoms. Many computations reported only consider one or a handful of structural models of mostly less than 100 atoms per cell. If only one structural model of this size is used, it is easy to miss important features or overemphasize others. The only way forward is to generate a series of wel-optimized structural models and assure that the ensemble is sufficiently large such that the computed observables assume reasonable distributions.

A second important problem is the misinterpretation of the artificially-introduced periodicity in the super cell approach. Many computations on amorphous materials apply a $\mathbf{k}$-point mesh to sample the first Brillouin zone, construct a band structure, and even use this to derive a band effective mass. Although this is the correct method for a crystalline material, in an amorphous phase, with the translational symmetry broken, the Bloch theorem does not hold, and this method is invalid.

A third problem concerns the creation of 'defects' without sufficient structural optimization. In a crystalline material the concept of a defect is very clear: atoms can either be missing from the lattice or additional atoms can be present at or in between lattice sites. In an amorphous material this entire concept disappears because there is no longer a lattice. Defects in amorphous materials are defined by either the presence of chemical bonds that are not present in the pristine crystalline state of the material or a deviation from the global stoichiometry. Just removing atoms from a previously-generated amorphous structural model may lead to local defect structures that are completely irrelevant. In most cases the structure will get stuck in a meta stable configuration, which may not survive under standard conditions. These problems often correlate with the lack of sufficient statistical sampling of the defect studied. 
 %%%

% maybe we should not rant about this here.... 
% A final point to mention here, not related to DFT calcuations in the misinterpretation of O1s XPS spectra in oxide semi conductors. 

Fortunately, several studies for the Zn-Sn-O system have been carried out according to best practice. In DFT computations of both crystalline and amorphous structures sub-gap states have been observed.\cite{Krner2012} A broad defect band above the valence edge is attributed to under-coordinated single oxygen atoms and a narrow band below the conduction edge to miscoordinated Sn-O complexes.\cite{Krner2012, Husein2020} In further work, including of-stoichiometric oxygen atomic ratios, the sub-gap states are shown to decrease at increasing oxygen content.\cite{Krner2014, Rucavado2017} In addition, systematic studies have shown that hybrid exchange correlation functionals PBE0, HSE06, or B3LYP agree with a self-interaction corrected (SIC) local-density-approximation (LDA) functional on occupied defect levels when similar treatments of the self-interaction are considered. However, for unoccupied levels, the hybrid functionals and the SIC approach lead to very different predictions.\cite{Krner2014_prb, Liu2016}

A computational Gibbs free energy evaluation on the Zn-Sn-O system is presented in Ref.~\citenum{Lee2013}. In agreement with experiment, the computations show that at medium temperature and pressure conditions a mixture of \ch{Zn2SnO4}, \ch{ZnO} and \ch{SnO2} is expected to form, while high pressure is needed to form \ch{ZnSnO3} below 1600K. These computations show a better agreement with experiment using the LDA functional than with the semi-local PBE  functional. The thermodynamics of ALD deposition of Zn-Sn-O systems has also been investigated at a DFT level of theory.\cite{Tanskanen2014}

In this work, we address gaps in the current state of knowledge by performing a systematic set of first principles computations. We first compare the four most stable Zn-Sn ratio oxides in both the crystalline and amorphous phases. We show that the most important electronic properties, electronic gap and electronic state overlap, do not vary much with the metal ratio. Next, we closely examine the effects of two types of impurities that impact the electronic properties in metal oxide semiconductors like IGZO. 

We investigate deviations from the stoichiometric amount of oxygen as various parameters in the deposition and integration of materials potentially impact the final deposited layer. This especially concerns oxides made of weakly oxidizing metals such as zinc and tin. The type of oxidizer during atomic layer deposition, the oxygen flow during physical vapor deposition, the choice of contact metals, and the forming gas anneal that is ubiquitous in silicon CMOS-compatible devices can all impact the final oxygen content.\cite{Medvedeva2017,Kim2014, Vogt2020, deJamblinnedeMeux2017, Kang2021, vanSetten2021_odef_igzo, Subhechha21_oxygen_tunnel, Kong2024, Guo2024} A sub-stoichiometric amount of oxygen is known of be a source of n-type doping in IGZO. Here we show that the oxygen binding for all Zn-Sn ratios is less stable than in 1:1:1 IGZO. It varies marginally with the Zn-Sn ratio with the material becoming less stable with increasing Sn content. 

Another source of n-type doping in IGZO is interstitial hydrogen. In this work we show that the Zn-Sn-O system behaves very similarly to IGZO in that aspect. Again, we show that a statistically correct sampling of of the computed observables over structural models is crucial to obtain meaningful numbers.

\section{Methodology}\label{method}

All computations reported in this paper are performed using the CP2K software package version 8.2.\cite{Khne2020} The hybrid Gaussian and plane wave density functional scheme of CP2K\cite{LIPPERT1997, Frigo2005, VandeVondele2005, Hutter2013, Bortnik2014} ensures that the dimension of the systems needed to reach low concentrations of defects are computationally feasible. We used the PBEsol generalized gradient approximation for the exchange correlation functional.\cite{Perdew1996, Perdew2008} The standard double $\zeta$ valence plus polarization (DZVP) basis sets\cite{VandeVondele2007} and pseudo potentials\cite{Krack2005, Hartwigsen1998, Goedecker1996} provided with CP2K are used. All calculations are performed using a single {\bf k}-point ($\Gamma$), to prevent effects caused by the artificial periodicity from entering the results. For the structure optimization, we use a maximum geometry change convergence criterion of 5~mBohr and a force convergence criterion of 1~m$E_H$/Bohr. We use a target accuracy for the electronic self-consistency convergence of $1\times10^{-6}$~$E_\mathrm{H}$. The preparation, execution, monitoring, and post-processing of the over 1250 computations reported in this work have been facilitated by our in-house Python package.

We base our structural modeling of the amorphous phase on crystalline prototype structures. These are chosen, with the help of the materials project database,\cite{Jain2013} as the most stable structure for the given metal ratio, see Tab.~\ref{tab:proto} for details. 

\begin{table}[!htb]
\centering
\begin{tabular}{llll}
\hline
\hline
material     & mp-id & $E_\mathrm{hull}$ & spacegroup \\
             &  & (eV/atom) &  \\
\hline
\ch{Zn2SnO2} & 35493   & 0.019 & orthorhombic Imma\\
\ch{ZnSnO3}  & 13334   & 0.041 & trigonal R3c\\
\ch{ZnSn2O5} & 1043000 & 0.063 & orthorhombic Cmcm\\
\ch{ZnSn3O7} & 1380234 & 0.238 & orthorhombic Pnma\\
\hline
\hline
\end{tabular}
\caption{Summary of the crystalline prototype structures used for the Zn-Sn-O system. The mp-id refers to the materials project database id and $E_\mathrm{hull}$ to the energy above the convex hull of this material. A positive value indicates that a reduction of the total energy could be obtained by phase segregation.}
\label{tab:proto}
\end{table}

The amorphous structural models used in this work are generated using the decorate and relax method proposed by Drabold et al.\cite{drabold09} In our experience, this approach leads to less defective structures at lower computational costs than melt and quench methodologies.\cite{deJamblinnedeMeux2018, vanSetten2021_odef_igzo, vanSetten2022screening} In each case we generate 20 super cell models of close to 200 atoms keeping the targeted stoichiometry. The relaxation uses a combination of the Broyden–Fletcher–Goldfarb–Shanno (BFGS) algorithm,\cite{BROYDEN1970, Fletcher1970, Goldfarb1970, Shanno1970} and time-stamped force-bias MonteCarlo (TFMC).\cite{Mees2012, Neyts2012} The details of the flow are provided in Tab.~\ref{tab:workflow}.

\begin{table}[!htb]
\centering
\begin{tabular}{llll}
\hline
\hline
step & method & steps & parameters \\
\hline
1 & BFGS & 20   & \\
2 & TFMC & 2000 & 300K\\
3 & BFGS & 2000 & \\
4 & TFMC & 2000 & 300K\\
5 & BFGS & 2000 & \\
6 & BFGS & 2000 & \\
\hline
\hline
\end{tabular}
\caption{Details on the optimization flow. The number of step in the BFGS is the sepcified maximum, which was never reached.}
\label{tab:workflow}
\end{table}

We use the Inverse State Weighted Overlap (ISWO) as a measure for electron and hole mobility to capture both the degree of delocalization of the electronic states and their spatial connectivity.\cite{deJamblinnedeMeux2018-iswo, vanSetten2022screening}

On all computed observables, gap, ISWO, cohesive energy, and reaction energies, we perform a statistical analysis to determine if a sufficient sampling of the amorphous phase has been achieved. For this we used the Shapiro-Wilk normality test.\cite{SHAPIRO1965} For the reaction energies we use a one-sided t-test to determine statistical significance.\cite{Student1908}

The stability with respect to oxygen scavenging is computed by comparing ensembles of stoichiometric structural models with ensembles of models containing one less oxygen atom. Note we do not remove oxygen atoms from previously-generated structures but generate new models containing less oxygen. Then a statistical analysis is performed to assess error bars and the statistical significance of the observed trends. 

Finally, hydrogen binding energies are computed with respect to a neutral gas phase molecule, i.e. \ce{H2}. We perform the studies by sampling all possible binding sites in a super cell containing 60 formula units of \ch{InGaZnO4} and 80 formula units of \ch{ZnSnO3} to reach a realistic hydrogen doping density of 0.2 at.\%. For IGZO we use a structure developed, in an identical manner, in previous work.\cite{Kruv2025} For \ch{ZnSnO3} we use a super cell of one of the amorphous models developed here that has no defects in the bandgap and create a $1\times1\times2$ super cell. Super cells of this size are needed to reach a realistic defect concentration. We first sample where an interstitial hydrogen atom can potentially bind. The sampling is based on a 3D grid with a spacing of 1~\AA. A box in this grid is used as a site if its center is not closer than 1.1~\AA\, of an existing atom. With this procedure we select a total of 124 sites for the crystalline phase and close to 1000 in the amorphous phase. In the crystalline phases only one replica of the original primitive cell is sampled inside of the full super cell.

For the hydrogen binding computations, we perform a direct 'local' minimization of the structure using the BFGS algorithm. This local optimization provides a clear picture of the energy distribution of the binding sites. We intentionally do not perform an {\em ab initio} TFMC or molecular dynamics simulated annealing type of optimization. This would collapse some of the more meta-stable binding sites into more stable ones. By performing a full, exhaustive, screening we will encounter the real global minimum by construction.

The, in total, over 1250 computations reported in this work (80 stoichiometic, 80 off stoichiometric, 975 H-doped amorphous, 124 H-doped crystal) are all set up, monitored, and processed using our in-house Python package.

\section{Results}

\subsection{Zn-Sn-O compositions with stoichiometric oxygen content}

\begin{figure}[!htb]
\centering
\includegraphics[width=0.8\columnwidth]{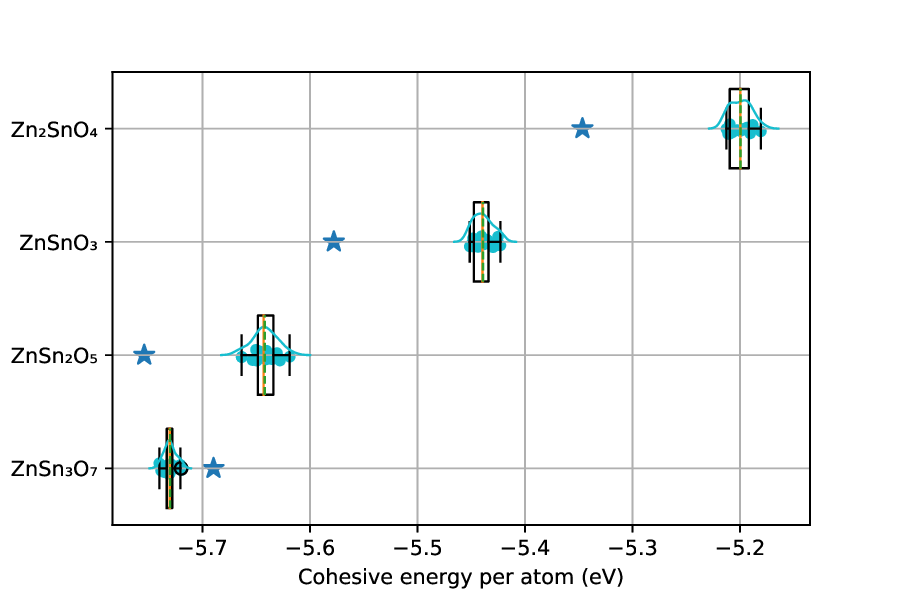}
\caption{Cohesive energy per atom (total number of atoms) of the crystalline and amorphous structural models of the Zn-Sn-O systems. The values of the crystal are indicated by the blue star, while the values of the amorphous structural models are visualized by the cyan dots, statistically summarized by a box plot and a kernel density estimate, cyan line. The solid orange and dashed green lines indicate the median and mean values, respectively. }
\label{fig:cepa}
\end{figure}

We first consider the fully oxidized, stoichiometric Zn$^{2+}$Sn$^{4+}$ materials, constructed such that for each Zn atom there is one oxygen atom and for each Sn atom there are two oxygen atoms. The pair correlation functions of the generated structures at the different steps of the optimization workflow are provided in the supplementary information (see Figs.~S1--S4). In the fully optimized amorphous models, we observe patterns that are very similar to those of the crystalline phase, with structural features preserved up to 12~\AA. Furthermore, no M--M or O--O defects are observed. The average densities of the generated models range from 6.0 (\ch{Zn2SnO4}) to 6.3 g/cm$^3$ (\ch{ZnSn3O7}). These densities are consistent with those required for amorphous \ch{SnZn4O}$_x$ to reach the regime in which the electrical properties become stable and favorable. The cohesive energies of the amorphous structural models are compared to those of the corresponding crystalline phases in Fig.~\ref{fig:cepa}.

Overall, we observe a narrow symmetric distribution of the cohesive energies of the amorphous structural models with only very few outliers. Interestingly, we observe that for the Zn:Sn 1:3 ratio the amorphous models are more stable than the crystalline structure. This agrees with the positive hull energy of the crystal, which indicates that this phase is metastable against phase segregation. It is, however, the most stable crystal crystal structure known for this ratio. A follow up study could be dedicated to search for lower energy crystalline structures.

\begin{figure}[!htb]
\centering
\includegraphics[width=0.8\columnwidth]{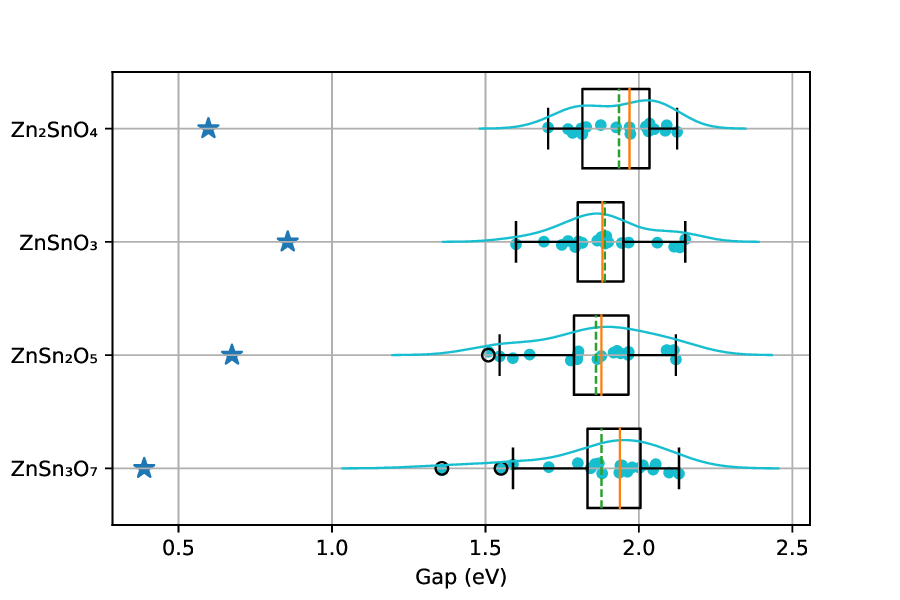}
\caption{The PBE bandgaps of the crystalline and amorphous structural models of the Zn-Sn-O system. The values of the crystal are indicated by the blue star, while the values of the amorphous structural models are visualized by the cyan dots, statistically summarized by a box plot and a kernel density estimate, cyan line. The solid orange and dashed green lines indicate the median and mean values, respectively.}
\label{fig:tg}
\end{figure}

The computed bandgaps, at PBEsol level, for the crystalline and amorphous structures of the Zn-Sn-O system are shown in Fig.~\ref{fig:tg}. The gaps of the crystalline materials are all about 0.2 eV smaller than those reported by the materials project.\cite{Jain2013} The relative trend, however, is conserved. This difference can easily be attributed to the differences in functional, PBE v.s. PBEsol, and pseudo potentials. The LDA gaps reported by K\"{o}rner {\em et all.}\cite{Krner2013} are, however, more than 50\% larger. This difference seems too large to be caused by the difference in functional, see Tab.~\ref{tab:gaps}. 

\begin{table}[!htb]
\centering
\begin{tabular}{llll}
\hline
\hline
material     & mp-id & $E_\mathrm{gap}$ (PBE) & $E_\mathrm{gap}$ (LDA SIC) \\
             &  & MP\cite{Jain2013}  & K\"{o}rner \em{et all.}\cite{Krner2013} \\
\hline
\ch{Zn2SnO2} & 35493   & 1.52 & 0.83  \\
\ch{ZnSnO3}  & 13334   & 1.47 & 1.08  \\
\ch{ZnSn2O5} & 1043000 & 1.40 & 0.85  \\
\ch{ZnSn3O7} & 1380234 & na   & 0.58  \\
\hline
\hline
\end{tabular}
\caption{Literature reported bandgaps for the crystalline phases of the Zn-Sn-O systems.}
\label{tab:gaps}
\end{table}

% pm gaps     Zn2SnO4 0.83, ZnSnO3 1.08, ZnSn2 0,85, ZnSn3 0.58
% korner lda  Zn2SnO4 1.52, ZnSnO3 1.47, ZnSn2 1.40, ZnSn3 0.58

For the amorphous phases, determining the relevant gap is less straightforward than for the crystalline ones. To obtain a value that can reasonably be compared to the macroscopic limit, a distribution over various structural models needs to be considered and for these models the defect states within the gap need to be disregarded.\cite{vanSetten2022screening}

For the bandgap distributions obtained in this way for the amorphous models, we observe an increase with respect to the crystalline values. The distributions of the gaps for the three compositions in the amorphous phase are very similar. Clearly, any trend that may be apparent in the minimal, maximal, average, or median values, is not significant. This rather similar value of the gaps at composition variation is a beneficial property from a device point of view. Based on these results, realistic variations in the composition of a macroscopic sample are not likely to cause large variations in the gap. A more uniform gap reduces Coulomb scattering\cite{Lordi2010} and the height and density of potential barriers in fluctuation induced tunneling.\cite{Sheng1980} This reduced scattering increases mobility due to an increased mean free time. 

Reported experimental bandgaps for amorphous zinc–tin oxides, obtained predominantly from optical measurements, range from 3 to 3.3~eV.\cite{thesis_son_2019, thesis_baran, Mullings2014} The discrepancy between these values and the computed gaps reported here (1.8–2~eV) is expected and reflects an intrinsic inconsistency of interpreting the Kohn–Sham eigenvalue gap as either an optical or a fundamental band gap, defined as the difference between the ionization energy and electron affinity.\cite{Baerends2013, vanSetten2017} In general, density functional theory systematically underestimates experimental fundamental band gaps, often by up to a factor of two, while typically preserving qualitative trends across materials.

The second relevant quantity concerning mobility is the effective mass of the charge carriers. In a crystalline material, where we can assume that transport is mainly occurring as band transport, the effective mass is related to the curvature of the electronic band structure. Applying this to an amorphous material is tempting but conceptually very wrong for two reasons. In amorphous materials, transport has more of a hopping nature and the entire concepts of electronic dispersion bands is invalid. The emergence of a band structure is the result of the Bloch theorem, which relies on translational symmetry. Since in the amorphous phase this symmetry is broken the concept of a band structure is incorrect. Although deriving an effective mass from the band structure curvature is erroneous, the concept of an effective mass is still valid. 

Any approach to quantify the hopping rate or effective mass in amorphous material needs to include in some way how the electronic states are spatially distributed and how close they are in energy. Both of these elements are contained in the concept of the Inverse State Weighted Overlap (ISWO).\cite{deJamblinnedeMeux2018-iswo, vanSetten2022screening} The state weighted overlap measures the state connection, and the inverse helps in the interpretation by making it behave as an effective mass: a low value leads to a high mobility.

\begin{figure}[!htb]
\centering
\includegraphics[width=0.8\columnwidth]{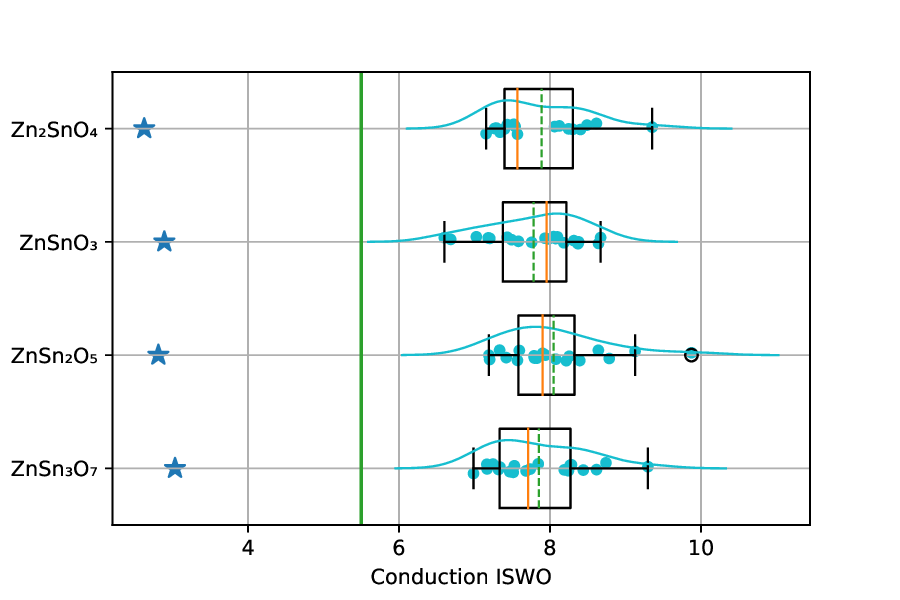}
\caption{The conduction ISWO values for the crystalline and amorphous structural models of the Zn-Sn-O system. The values of the crystal are indicated by the blue star, while the values of the amorphous structural models are visualized by the cyan dots, statistically summarized by a box plot and a kernel density estimate, cyan line. The solid orange and dashed green lines indicate the median and mean values, respectively. For reference, the mean value of the archetypical amorphous oxide semiconductor \ch{InGaZnO4} is given by the green vertical.}
\label{fig:iswo_c}
\end{figure}

The distributions of the mean ISWO values in the lowest 0.3eV of the conduction states for the four Zn-Sn-O compositions are shown in Fig.~\ref{fig:iswo_c}. These distributions for the amorphous models are compared to the values calculated for the crystalline phase, which are indicated by the star symbols. Since ISWO is a dimensionless quantity, we also provide a reference in the average ISWO conduction values of IGZO. We observe that the results for the crystals seem very promising; the values are all well below the average value of amorphous IGZO. Unfortunately, the transition to the amorphous phase causes a severe drop in predicted mobility compared to those of the crystalline phase. In the amorphous models, no significant trend is present as a function of the Zn-Sn ratio.

\begin{figure}[!htb]
\centering
\includegraphics[width=0.8\columnwidth]{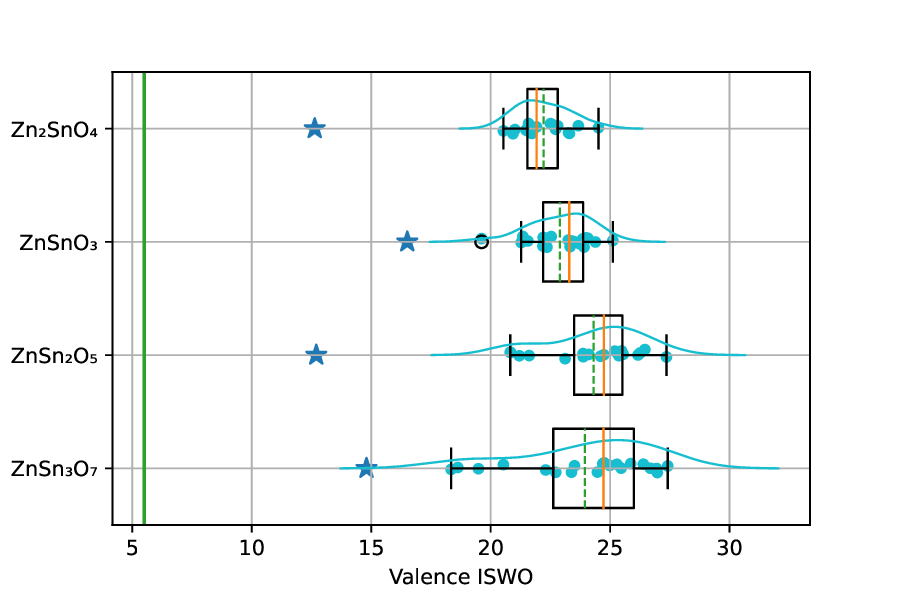}
\caption{The valence ISWO values for the crystalline and amorphous structural models of the Zn-Sn-O system. The values of the crystal are indicated by the blue star, while the values of the amorphous structural models are visualized by the cyan dots, statistically summarized by a box plot and a kernel density estimate, cyan line. The solid orange and dashed green lines indicate the median and mean values, respectively. For reference, the mean conduction ISWO value of the archetypical oxide semiconductor amorphous IGZO is given by the green vertical.}
\label{fig:iswo_v}
\end{figure}

In Fig.~\ref{fig:iswo_v}, we also consider the ISWO values of the top of the valence states. These values are a measure for how mobile holes would be in the materials. The green line again indicates the value for electrons in IGZO. The distributions around 20 to 25 show that we can expect very low hole mobility. This agrees with the observation that the upper part of the valence states consists of a hybridization of Zn-d and O-p atomic orbitals. For a n-type transistor, this would mean a very low off-current can be expected. It also means that, if p-type doping could be achieved, p-type mobility would be very low.

\subsection{Substoichiometric oxygen content}

In a crystal, a vacancy is a well defined concept, it is just an empty lattice site. In the amorphous phase, where no lattice is present, the definition is not very clear. One could start from an atomistic model of the amorphous phase and take out one atom and call this a vacancy. However, due to the flexible nature of the amorphous network, the material will adjust through plastic deformation around the created 'vacancy', minimizing the energy. Any proper global energy minimization process will cause the original vacancy to be, in a sense, delocalized, and we are just left with a global stoichiometry disturbance but no detectable point defect. A more systematic approach, hence, consists in generating proper statistics on stoichiometric and off-stoichiometric samples until statistically significant differences in energy distributions are obtained.

\begin{figure}[!htb]
\centering
\includegraphics[width=0.8\columnwidth]{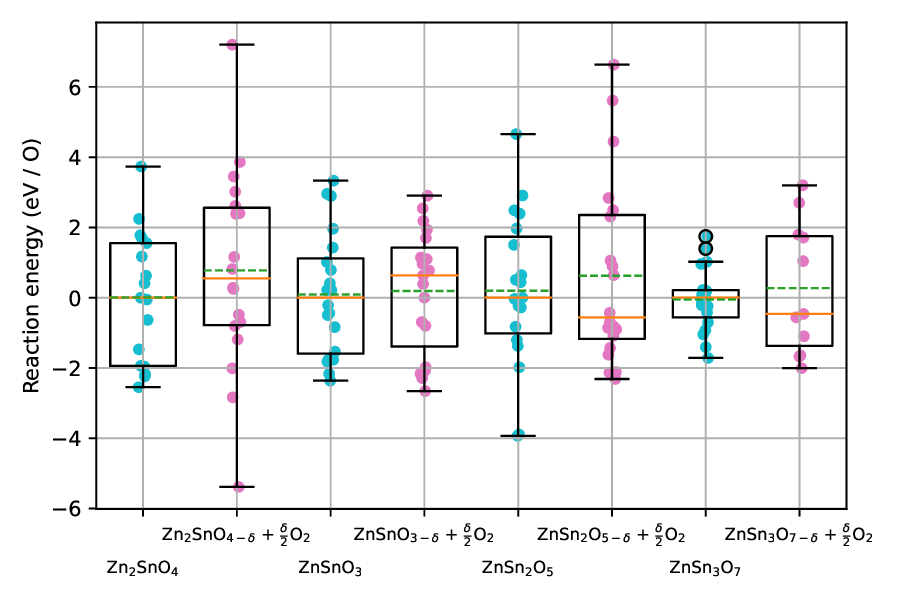}
\caption{The computed distributions of oxygen desorption reactions for \ch{Zn2SnO3}, \ch{ZnSnO4}, and \ch{ZnSn2O5} at 0.8\% oxygen release. For each material, the median of the energies of the stoichiometric models is set at 0 and the energies of the corresponding oxygen reduced models are plotted relative to that. In the boxplots the solid orange and dashed green lines indicate the median and mean values, respectively. The reference chemical potential for oxygen is taken at half an oxygen molecule at 0K.}
\label{fig:o_desorpb_amorph}
\end{figure}

Figure~\ref{fig:o_desorpb_amorph} shows the distribution of the electronic total energies for the amorphous models. For each Zn-Sn ratio, the median value of the stoichiometric compositions is set to 0 and for the off-stoichiometric version the amount of missing oxygen atom is compensated by half an oxygen molecule. For example, the median values for \ch{Zn2SnO4} indicate that it takes about 0.5~eV to remove on oxygen atom with respect to forming di-oxygen gas. This number is significantly lower than what we previously calculated for the lowest oxygen defect concentration for IGZO.\cite{vanSetten2021_odef_igzo}  From Fig.~\ref{fig:o_desorpb_amorph}, we also observe that with increasing Sn content, the oxygen seems to be bound more weakly. This is, however, not a statistically significant trend. The p-values of a t-test between the stoichiometric and oxygen reduced models are all too large for the distributions to be significantly different. Moreover, for all compositions the oxygen binding is weaker than the formation of \ch{H2O} (241.8 kJ/mol 2.58 eV/molecule)\cite{CCODATA}. This indicates that all compositions are vulnerable to oxygen scavenging by \ch{H2} gas exposure. When calculating the distributions of only ten amorphous models per composition, no significantly different distributions of energies are obtained between the stoichiometric and oxygen deficient model are obtained, the t-test p-values range from 0.1 to 0.3. Moving to twenty models per composition lessens this problem. The reaction energies extracted for the results in Fig.~\ref{fig:o_desorpb_amorph} are summarized in Tab.~\ref{tab:reaction_energies}.

The oxygen binding values calculated in this way resemble a chemical potential like quantity. In this sense, they represent an upper limit of a true bulk amorphous phase. In realistic semiconductor devices, layers are only several nm thick. The effects of this finite thickness and the related interfaces are not included in the current results. 

\begin{table}[!htb]
\centering
\begin{tabular}{lrrr}
\hline
\hline
reaction      & mean $\Delta E$& median $\Delta E$ & p-value \\
&(eV / O)&(eV / O)&\\
\hline
Zn$_2$SnO$_4$ $\rightarrow$ Zn$_2$SnO$_{4-\delta}$ + $\frac{\delta}{2}$O$_2$ & 0.770 & 0.550 & 0.358 \\
ZnSnO$_3$ $\rightarrow$ ZnSnO$_{3-\delta}$ + $\frac{\delta}{2}$O$_2$ & 0.100 & 0.638 & 0.861 \\
ZnSn$_2$O$_5$ $\rightarrow$ ZnSn$_2$O$_{5-\delta}$ + $\frac{\delta}{2}$O$_2$ & 0.428 & -0.560 & 0.587 \\
ZnSn$_3$O$_7$ $\rightarrow$ ZnSn$_3$O$_{7-\delta}$ + $\frac{\delta}{2}$O$_2$ & 0.324 & -0.462 & 0.527\\
\hline
\hline
\end{tabular}
\caption{Extracted reaction energies. The mean $\Delta$E and median $\Delta$E represent reaction energies based on the differences between the means and medians of the distributions of the stoichiometric and oxygen reduced models, respectively. The p-value is calculated for a two-sided t-test.}
\label{tab:reaction_energies}
\end{table}

\begin{figure}[!htb]
\centering
\includegraphics[width=0.8\columnwidth]{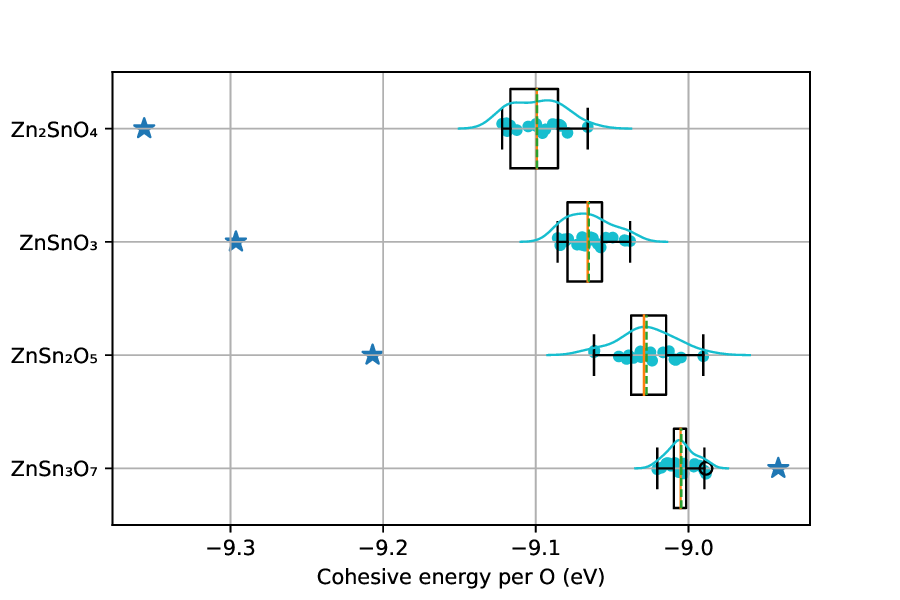}
\caption{Cohesive energy per atom of the crystalline and amorphous structural models of the Zn-Sn-O system. The values of the crystal are indicated by a blue star, the values of the amorphous structural models are visualized by a box plot, the solid orange and dashed green lines indicate the median and mean values, respectively.}
\label{fig:cepo}
\end{figure}

In Fig.~\ref{fig:cepo} we provide, for comparison, the cohesive energies of the compositions per oxygen. Renormalized like this, in contrast to the per atom data shown in Fig.~\ref{fig:cepa}, we see that the same trends hold.

\subsection{Interstitial hydrogen}

\begin{figure}[!htb]
\centering
\includegraphics[width=0.6\columnwidth]{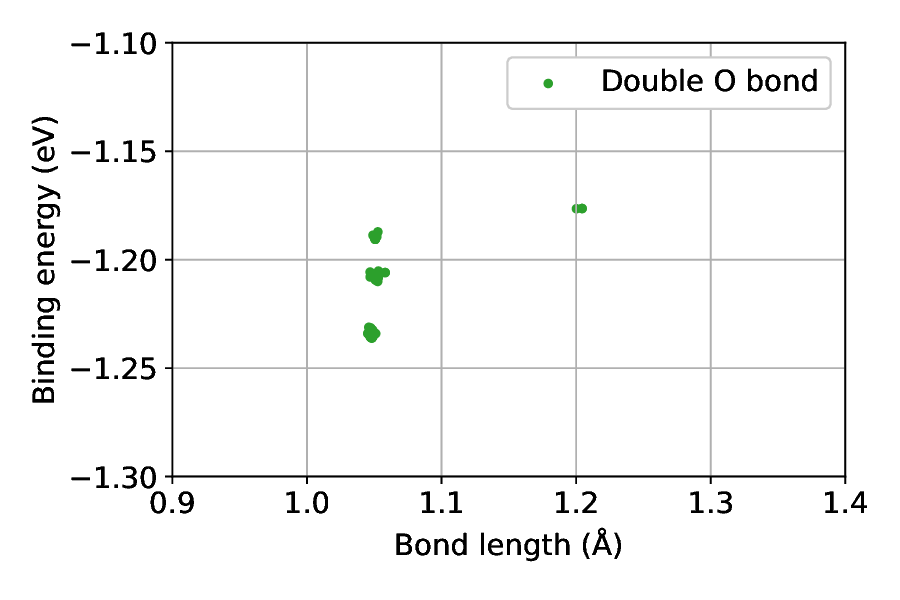}
\caption{The computed distribution of hydrogen binding energy and nearest neighbour distance and type in \ch{ZnSnO3}. Negative values indicate stable binding.}
\label{fig:h_binding_crystal}
\end{figure}

Finally, we consider the direct interaction of the Zn–Sn–O system with hydrogen. Due to the large number of computations involved, we restrict our analysis to the 1:1 Zn:Sn ratio. We use close to 400 atom models for this to ensure accurate results. Since the overall cohesive energy does not change drastically with the Zn:Sn ratio (see Fig.~\ref{fig:cepa} and \ref{fig:cepo}) we estimate that the defect state results can be extended to other Zn:Sn ratios. However, as the ratios that optimize the measured field-effect mobility tend to lie on the Zn-rich side,\cite{Allemang2020} further investigations may be needed to understand the impact of hydrogen on materials with other ZN-Sn ratios. In Fig.~\ref{fig:h_binding_crystal}, we first examine hydrogen binding in the crystalline phase. We observe four distinct binding sites, indicating that even in the crystalline phase the situation can be rather complex. As in the case of IGZO,\cite{vanSetten2024_igzo_h} we find that hydrogen preferentially binds to oxygen and that this binding is stable with respect to \ch{H2} in the gas phase.

\begin{figure}[!htb]
\centering
\includegraphics[width=0.8\columnwidth]{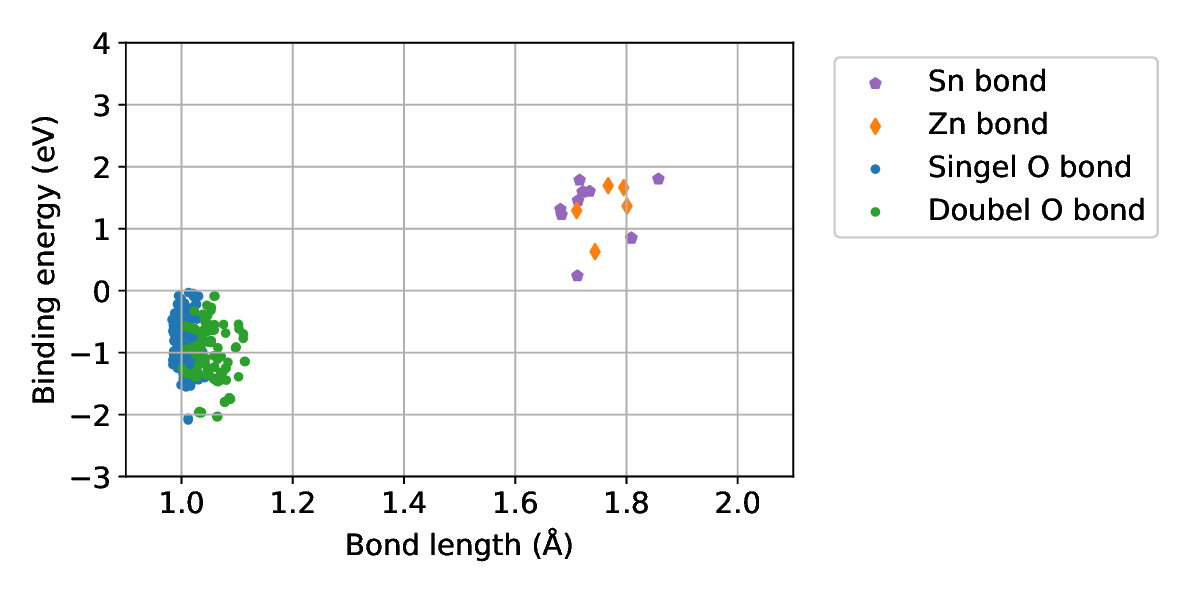}
\caption{The computed distribution of hydrogen binding energy and nearest neighbour distance and type in amorphous \ch{ZnSnO3}. Negative values indicate stable binding.}
\label{fig:h_binding_amorph_znsno}
\end{figure}

\begin{figure}[!htb]
\centering
\includegraphics[width=0.8\columnwidth]{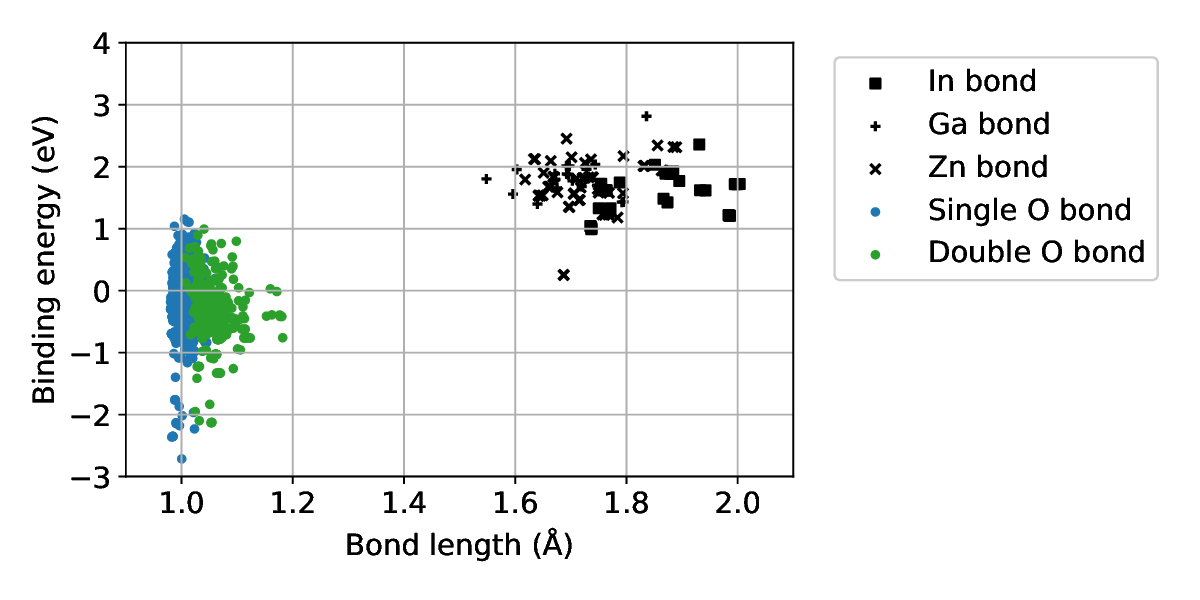}
\caption{The computed distribution of hydrogen binding energy and nearest neighbour distance and type in amorphous \ch{InGaZnO4}. Negative values indicate stable binding.}
\label{fig:h_binding_amorph_igzo}
\end{figure}

Next, we switch to the amorphous phase in Fig.~\ref{fig:h_binding_amorph_znsno}, where negative binding energy means stable binding with respect to hydrogen gas. Three types of hydrogen binding are present. The blue dots indicate hydrogen binding to a single oxygen atom. The green dots indicate binding to oxygen but here a second oxygen atom is also present forming a hydrogen bond. Finally, we have the purple and orange symbols for hydrogen binding to one of the metals. The results show that hydrogen binding to oxygen is stable while binding to the metal atoms is metastable. The situation is qualitatively similar to what we observe in IGZO, shown for reference in Fig.~\ref{fig:h_binding_amorph_igzo}.\cite{Kruv2023,Kruv2025}

\section{Discussion on atomistic modeling of amorphous phases}

To study amorphous materials with first-principles methods, atomistic structural models are first needed. The reliability of any computed quantity ultimately depends on the quality of these models. In crystalline phases, periodicity can be exploited to compute macroscopic properties and quantities such as lattice parameters, and X-ray diffraction patterns can be used to validate the quality of the structure.

In the amorphous phase, the supercell approach can be used to suppress finite-size effects and mimic a macroscopic sample, but convergence with respect to the cell size is needed to prevent artificial periodicity from impacting the calculated values of the observables. In addition, sufficient sampling of different models is needed to obtain an indication of the confidence interval of the calculated quantity of interest and to make claims about significant differences between materials. For a given material, with its specific bond energies and a given level of structural optimization, one can expect that a sufficiently large sampling of structures will lead to a normal distribution of that observable. How close the distribution is to a Gaussian is a measure of the quality of the sampling. The width of the distribution is determined by the structural variation and the sensitivity of the observable to that variation. In this work, we showed that, for the optimization scheme used here, a sample size of 20 models per composition and about 200 atoms per model leads to distributions of the total energies, electronic observables, and pair correlation functions that are useful for comparing different materials and that begin to mimic true macroscopic distributions. We thus propose that best practice for first-principles computations of amorphous materials includes the use of a sufficiently large sample size and an estimation of the statistical distribution of the calculated observables.

Studying missing atoms or impurities in amorphous materials is much more challenging than in the crystalline phase. Due to the large number of degrees of freedom, an amorphous material can adjust to a missing atom to a much larger extent. At macroscopic time scales under standard conditions, the material has sufficient time to eliminate any local point-like defect. Thus, careful sampling and optimization are crucial. In this work, we compare distributions of stoichiometric models and models with one oxygen atom removed. These distributions are difficult to converge. Interstitial atoms, in our case hydrogen, are also not trivial to study in amorphous materials. In our results, we see that only exhaustive sampling, i.e., 1000 potential sites for \ch{ZnSnO3}, provides a complete picture. Current modeling of amorphous materials often lacks sufficient sampling to obtain reliable results.
 
\section{Conclusions for the Zn-Sn-O system}

%4*20 = 80 stoichiometic
%4*20 = 80 of stoichiometric
%975 h-doped amorphous
%124 h-doped crystal

In this work, we systematically studied the Zn-Sn-O system in the amorphous phase. In total, over 1250 atomistic models containing 200 atoms (for structural, electronic, and oxygen off-stoichiometric modeling) and 400 atoms (for hydrogen doping) were evaluated. Using this number of models, we obtain distributions that begin to approach normal distributions and can be interpreted as macroscopic results for all observables considered.

The conduction ISWO values, which form a proxy for the electron effective mass, in the Zn-Sn-O compositions are less favorable than those of IGZO. Their variation as a function of metal ratio, however, is smaller.\cite{Kruv2023} Compositional variation in macroscopic samples is thus expected to lead to smaller variation in the observables. Especially for the bandgap this is favorable. Smaller gap variation should reduce Coulombic scattering and hence increase the electron mean free time. 

The stability against oxygen scavenging and hydrogen doping of the Zn-Sn-O system are qualitatively similar to those of IGZO. It is hence not clear if this material can be the solution to the electrical instabilities observed in IGZO based transistors.

Taken the above all together, we conclude that from a performance point of view the Zn-Sn-O system could be a reasonable alternative for IGZO but from a reliability point of view the hydrogen migration related problems in IGZO may be present as well.

\bibliography{bibtex}

@book{CCODATA,
    author = {J. D. Cox and D. D. Wagman and V. A. Medvedev}, 
    title = {CCODATA Key Values for Thermodynamics},
    publisher = {Hemisphere Publishing Corp.},
    year = 1984
}

@phdthesis{thesis_baran,
  doi = {10.7302/8611},
  url = {http://deepblue.lib.umich.edu/handle/2027.42/178154},
  author = {Demir,  Baran},
  language = {en},
  title = {Solution-Based Chemical Deposition of Wide and Ultra-Wide Bandgap Oxide Semiconductors},
  publisher = {University of Michigan},
  year = {2023}
}

@phdthesis{thesis_son_2019,
  url = {https://hdl.handle.net/2027.42/149911},
  author = {Son, Youngbae },
  language = {en},
  title = {Solution-Processed Amorphous Oxide Semiconductors for Thin-Film Power Management Circuitry },
  publisher = {University of Michigan},
  year = {2019}
}

@article{Mullings2014,
  title = {Thin film characterization of zinc tin oxide deposited by thermal atomic layer deposition},
  volume = {556},
  ISSN = {0040-6090},
  url = {http://dx.doi.org/10.1016/j.tsf.2014.01.068},
  DOI = {10.1016/j.tsf.2014.01.068},
  journal = {Thin Solid Films},
  publisher = {Elsevier BV},
  author = {Mullings,  Marja N. and H\"{a}gglund,  Carl and Tanskanen,  Jukka T. and Yee,  Yesheng and Geyer,  Scott and Bent,  Stacey F.},
  year = {2014},
  month = apr,
  pages = {186–194}
}

@misc{endangered_elements,
  title = {Endangered Elements},
  howpublished = {\url{https://www.acs.org/green-chemistry-sustainability/research-innovation/endangered-elements.html}},  note = {accessed 2025-01-23}
}

@article{Kruv2025,
  title = {In-Poor IGZO: Superior Resilience to Hydrogen in Forming Gas Anneal and PBTI},
  volume = {7},
  ISSN = {2637-6113},
  url = {http://dx.doi.org/10.1021/acsaelm.5c00383},
  DOI = {10.1021/acsaelm.5c00383},
  number = {9},
  journal = {ACS Appl. Electron. Mater.},
  publisher = {American Chemical Society (ACS)},
  author = {Kruv,  Anastasiia and {van Setten},  Michiel J. and Chasin,  Adrian and Matsubayashi,  Daisuke and Dekkers,  Hendrik F. W. and Pavel,  Alexandru and Wan,  Yiqun and Trivedi,  Kruti and Rassoul,  Nouredine and Li,  Jie and Jiang,  Yuchao and Subhechha,  Subhali and Pourtois,  Geoffrey and Belmonte,  Attilio and Kar,  Gouri Sankar},
  year = {2025},
  month = apr,
  pages = {4210–4219}
}

@article{Xu2023,
  title = {Advances in data‐assisted high‐throughput computations for material design},
  volume = {1},
  ISSN = {2940-9497},
  url = {http://dx.doi.org/10.1002/mgea.11},
  DOI = {10.1002/mgea.11},
  number = {1},
  journal = {Materials Genome Engineering Advances},
  publisher = {Wiley},
  author = {Xu,  Dingguo and Zhang,  Qiao and Huo,  Xiangyu and Wang,  Yitong and Yang,  Mingli},
  year = {2023},
  month = sep 
}

@article{Mees2012,
  title = {Uniform-acceptance force-bias Monte Carlo method with time scale to study solid-state diffusion},
  volume = {85},
  ISSN = {1550-235X},
  url = {http://dx.doi.org/10.1103/PHYSREVB.85.134301},
  DOI = {10.1103/physrevb.85.134301},
  pages = {134301},
  number = {13},
  journal = {Phys. Rev. B},
  publisher = {American Physical Society (APS)},
  author = {Mees,  Maarten J. and Pourtois,  Geoffrey and Neyts,  Erik C. and Thijsse,  Barend J. and Stesmans,  André},
  year = {2012},
  month = apr 
}

@article{Neyts2012,
  title = {Combining molecular dynamics with Monte Carlo simulations: implementations and applications},
  volume = {132},
  ISSN = {1432-2234},
  url = {http://dx.doi.org/10.1007/s00214-012-1320-x},
  DOI = {10.1007/s00214-012-1320-x},
  number = {2},
  journal = {Theor. Chem. Acc.},
  publisher = {Springer Science and Business Media LLC},
  author = {Neyts,  Erik C. and Bogaerts,  Annemie},
  pages = {1320},
  year = {2014},
  month = dec 
}

@article{Fernandes2018,
  title = {A Sustainable Approach to Flexible Electronics with Zinc‐Tin Oxide Thin‐Film Transistors},
  volume = {4},
  ISSN = {2199-160X},
  url = {http://dx.doi.org/10.1002/aelm.201800032},
  DOI = {10.1002/aelm.201800032},
  number = {7},
  journal = {Adv. Electron. Mater.},
  pages = {1800032},
  publisher = {Wiley},
  author = {Fernandes,  Cristina and Santa,  Ana and Santos,  Angelo and Bahubalindruni,  Pydi and Deuermeier,  Jonas and Martins,  Rodrigo and Fortunato,  Elvira and Barquinha,  Pedro},
  year = {2018},
  month = may 
}

@article{Allemang2020,
  title = {High‐Performance Zinc Tin Oxide TFTs with Active Layers Deposited by Atomic Layer Deposition},
  volume = {6},
  ISSN = {2199-160X},
  url = {http://dx.doi.org/10.1002/aelm.202000195},
  DOI = {10.1002/aelm.202000195},
  number = {7},
  journal = {Adv. Electron. Mater.},
  publisher = {Wiley},
  author = {Allemang,  Christopher R. and Cho,  Tae H. and Trejo,  Orlando and Ravan,  Shantam and Rodríguez,  Robin E. and Dasgupta,  Neil P. and Peterson,  Rebecca L.},
  pages = {2000195},
  year = {2020},
  month = jun 
}

@article{Banger2010,
  title = {Low-temperature,  high-performance solution-processed metal oxide thin-film transistors formed by a ‘sol–gel on chip’ process},
  volume = {10},
  ISSN = {1476-4660},
  url = {http://dx.doi.org/10.1038/nmat2914},
  DOI = {10.1038/nmat2914},
  number = {1},
  journal = {Nat. Mater.},
  publisher = {Springer Science and Business Media LLC},
  author = {Banger,  K. K. and Yamashita,  Y. and Mori,  K. and Peterson,  R. L. and Leedham,  T. and Rickard,  J. and Sirringhaus,  H.},
  year = {2010},
  month = dec,
  pages = {45–50}
}

@article{Rinaudo2023,
  title = {Degradation Mapping and Impact of Device Dimension on IGZO TFTs BTI},
  volume = {23},
  ISSN = {1558-2574},
  url = {http://dx.doi.org/10.1109/TDMR.2023.3282298},
  DOI = {10.1109/tdmr.2023.3282298},
  number = {3},
  journal = {IEEE Trans. Device Mater. Reliab.},
  publisher = {Institute of Electrical and Electronics Engineers (IEEE)},
  author = {Rinaudo,  Pietro and Chasin,  A. and Franco,  J. and Wu,  Z. and Subhechha,  S. and Arutchelvan,  G. and Eneman,  G. and Ramana,  B. Y. V. and Rassoul,  N. and Delhougne,  R. and Kaczer,  B. and De Wolf,  I. and Kar,  G. S.},
  year = {2023},
  month = sep,
  pages = {337–345}
}

@article{Liu2024,
  title = {TFT-Based Near-Sensor In-Memory Computing: Circuits and Architecture Perspectives of Large-Area eDRAM and ROM CiM Chips},
  volume = {71},
  ISSN = {1558-0806},
  url = {http://dx.doi.org/10.1109/TCSI.2023.3340114},
  DOI = {10.1109/tcsi.2023.3340114},
  number = {2},
  journal = {IEEE Trans. Circuits Syst. I Regul. Pap.},
  publisher = {Institute of Electrical and Electronics Engineers (IEEE)},
  author = {Liu,  Jialong and Tang,  Wenjun and Li,  Hongtian and Chen,  Deyun and Long,  Weihang and Liu,  Yongpan and Jiang,  Chen and Yang,  Huazhong and Li,  Xueqing},
  year = {2024},
  month = feb,
  pages = {620–633}
}

@inproceedings{Wu2022,
  title = {Characterizing and Modelling of the BTI Reliability in IGZO-TFT using Light-assisted I-V Spectroscopy},
  url = {http://dx.doi.org/10.1109/IEDM45625.2022.10019454},
  DOI = {10.1109/iedm45625.2022.10019454},
  booktitle = {2022 International Electron Devices Meeting (IEDM)},
  publisher = {IEEE},
  author = {Wu,  Z. and Chasin,  A. and Franco,  J. and Subhechha,  S. and Dekkers,  H. and Bhuvaneshwari,  Y.V. and Belmonte,  A. and Rassoul,  N. and {van Setten},  M.J. and Afanas’Ev,  V. and Delhougne,  R. and Kaczer,  B. and Kar,  G.S.},
  year = {2022},
  month = dec 
}

@article{Krner2014,
  title = {Density-functional theory study of stability and subgap states of crystalline and amorphous Zn–Sn–O},
  volume = {555},
  ISSN = {0040-6090},
  url = {http://dx.doi.org/10.1016/j.tsf.2013.05.146},
  DOI = {10.1016/j.tsf.2013.05.146},
  journal = {Thin Solid Films},
  publisher = {Elsevier BV},
  author = {K\"{o}rner,  Wolfgang and Els\"{a}sser,  Christian},
  year = {2014},
  month = mar,
  pages = {81–86}
}

@article{Krner2014_prb,
  title = {Prediction of subgap states in Zn- and Sn-based oxides using various exchange-correlation functionals},
  volume = {90},
  ISSN = {1550-235X},
  url = {http://dx.doi.org/10.1103/PhysRevB.90.195142},
  DOI = {10.1103/physrevb.90.195142},
  pages = {195142},
  number = {19},
  journal = {Phys. Rev. B},
  publisher = {American Physical Society (APS)},
  author = {K\"{o}rner,  Wolfgang and Urban,  Daniel F. and Ramo,  David Muñoz and Bristowe,  Paul D. and Els\"{a}sser,  Christian},
  year = {2014},
  month = nov 
}

@article{Krner2012,
  title = {Analysis of electronic subgap states in amorphous semiconductor oxides based on the example of Zn-Sn-O systems},
  volume = {86},
  ISSN = {1550-235X},
  url = {http://dx.doi.org/10.1103/PhysRevB.86.165210},
  DOI = {10.1103/physrevb.86.165210},
  number = {16},
  journal = {Phys. Rev. B},
  pages = {165210},
  publisher = {American Physical Society (APS)},
  author = {K\"{o}rner,  Wolfgang and Gumbsch,  Peter and Els\"{a}sser,  Christian},
  year = {2012},
  month = oct 
}

@article{Walsh2009,
  title = {Interplay between Order and Disorder in the High Performance of Amorphous Transparent Conducting Oxides},
  volume = {21},
  ISSN = {1520-5002},
  url = {http://dx.doi.org/10.1021/cm9020113},
  DOI = {10.1021/cm9020113},
  number = {21},
  journal = {Chem. Mater.},
  publisher = {American Chemical Society (ACS)},
  author = {Walsh,  Aron and Da Silva,  Juarez L. F. and Wei,  Su-Huai},
  year = {2009},
  month = oct,
  pages = {5119–5124}
}

@article{Baerends2013,
  title = {The Kohn–Sham gap,  the fundamental gap and the optical gap: the physical meaning of occupied and virtual Kohn–Sham orbital energies},
  volume = {15},
  ISSN = {1463-9084},
  url = {http://dx.doi.org/10.1039/C3CP52547C},
  DOI = {10.1039/c3cp52547c},
  number = {39},
  journal = {Physical Chemistry Chemical Physics},
  publisher = {Royal Society of Chemistry (RSC)},
  author = {Baerends,  E. J. and Gritsenko,  O. V. and van Meer,  R.},
  year = {2013},
  pages = {16408}
}

@article{Powell1989,
  title = {The physics of amorphous-silicon thin-film transistors},
  volume = {36},
  ISSN = {0018-9383},
  url = {http://dx.doi.org/10.1109/16.40933},
  DOI = {10.1109/16.40933},
  number = {12},
  journal = {IEEE Trans. Electron Devices},
  publisher = {Institute of Electrical and Electronics Engineers (IEEE)},
  author = {Powell,  M.J.},
  year = {1989},
  pages = {2753–2763}
}

@article{Student1908,
  title = {The Probable Error of a Mean},
  volume = {6},
  ISSN = {0006-3444},
  url = {http://dx.doi.org/10.2307/2331554},
  DOI = {10.2307/2331554},
  number = {1},
  journal = {Biometrika},
  publisher = {JSTOR},
  author = {Student},
  year = {1908},
  month = mar,
  pages = {1}
}

@article{Rucavado2017,
  title = {Enhancing the optoelectronic properties of amorphous zinc tin oxide by subgap defect passivation: A theoretical and experimental demonstration},
  volume = {95},
  ISSN = {2469-9969},
  url = {http://dx.doi.org/10.1103/PhysRevB.95.245204},
  DOI = {10.1103/physrevb.95.245204},
  number = {24},
  journal = {Phys. Rev. B},
  pages = {245204},
  publisher = {American Physical Society (APS)},
  author = {Rucavado,  Esteban and Jeangros,  Quentin and Urban,  Daniel F. and Holovský,  Jakub and Remes,  Zdenek and Duchamp,  Martial and Landucci,  Federica and Dunin-Borkowski,  Rafal E. and K\"{o}rner,  Wolfgang and Els\"{a}sser,  Christian and Hessler-Wyser,  Aïcha and Morales-Masis,  Monica and Ballif,  Christophe},
  year = {2017},
  month = jun 
}

@article{Zeng2024,
  title = {Thermal Stability of Amorphous Metal Oxides: The Interplay of Secondary Cations,  Degree of Substitution,  and Local Structure},
  ISSN = {1520-5002},
  volume = {36},
  url = {http://dx.doi.org/10.1021/acs.chemmater.4c00351},
  DOI = {10.1021/acs.chemmater.4c00351},
  journal = {Chem. Mater.},
  publisher = {American Chemical Society (ACS)},
  author = {Zeng,  Li and Buchholz,  D. Bruce and Keane,  Denis T. and Marks,  Tobin J. and Medvedeva,  Julia E. and Bedzyk,  Michael J.},
  year = {2024},
  pages = {5965–5975},
  month = jun 
}

@article{Choi2024,
  title = {Review of Material Properties of Oxide Semiconductor Thin Films Grown by Atomic Layer Deposition for Next-Generation 3D Dynamic Random-Access Memory Devices},
  volume = {36},
  ISSN = {1520-5002},
  url = {http://dx.doi.org/10.1021/acs.chemmater.3c02223},
  DOI = {10.1021/acs.chemmater.3c02223},
  number = {5},
  journal = {Chem. Mater.},
  publisher = {American Chemical Society (ACS)},
  author = {Choi,  Ae Rim and Lim,  Dong Hyun and Shin,  So-Yeon and Kang,  Hye Joo and Kim,  Dohee and Kim,  Ja-Yong and Ahn,  Youngbae and Ryu,  Seung Wook and Oh,  Il-Kwon},
  year = {2024},
  month = feb,
  pages = {2194–2219}
}

@article{Nadarajah2015,
  title = {Amorphous In–Ga–Zn Oxide Semiconducting Thin Films with High Mobility from Electrochemically Generated Aqueous Nanocluster Inks},
  volume = {27},
  ISSN = {1520-5002},
  url = {http://dx.doi.org/10.1021/acs.chemmater.5b01813},
  DOI = {10.1021/acs.chemmater.5b01813},
  number = {16},
  journal = {Chem. Mater.},
  publisher = {American Chemical Society (ACS)},
  author = {Nadarajah,  Athavan and Wu,  Mahkah Z. B. and Archila,  Kevin and Kast,  Matthew G. and Smith,  Adam M. and Chiang,  Tsung H. and Keszler,  Douglas A. and Wager,  John F. and Boettcher,  Shannon W.},
  year = {2015},
  month = aug,
  pages = {5587–5596}
}

@article{Sil2022,
  title = {Fluoride Doping in Crystalline and Amorphous Indium Oxide Semiconductors},
  volume = {34},
  ISSN = {1520-5002},
  url = {http://dx.doi.org/10.1021/acs.chemmater.2c00053},
  DOI = {10.1021/acs.chemmater.2c00053},
  number = {7},
  journal = {Chem. Mater.},
  publisher = {American Chemical Society (ACS)},
  author = {Sil,  Aritra and Deck,  Michael J. and Goldfine,  Elise A. and Zhang,  Chi and Patel,  Sawankumar V. and Flynn,  Steven and Liu,  Haoyu and Chien,  Po-Hsiu and Poeppelmeier,  Kenneth R. and Dravid,  Vinayak P. and Bedzyk,  Michael J. and Medvedeva,  Julia E. and Hu,  Yan-Yan and Facchetti,  Antonio and Marks,  Tobin J.},
  year = {2022},
  month = mar,
  pages = {3253–3266}
}

@article{Buchholz2014,
  title = {The Structure and Properties of Amorphous Indium Oxide},
  volume = {26},
  ISSN = {1520-5002},
  url = {http://dx.doi.org/10.1021/cm502689x},
  DOI = {10.1021/cm502689x},
  number = {18},
  journal = {Chem. Mater.},
  publisher = {American Chemical Society (ACS)},
  author = {Buchholz,  D. Bruce and Ma,  Qing and Alducin,  Diego and Ponce,  Arturo and Jose-Yacaman,  Miguel and Khanal,  Rabi and Medvedeva,  Julia E. and Chang,  Robert P. H.},
  year = {2014},
  month = sep,
  pages = {5401–5411}
}

@article{Thyagarajan2020,
  title = {A Database of Porous Rigid Amorphous Materials},
  volume = {32},
  ISSN = {1520-5002},
  url = {http://dx.doi.org/10.1021/acs.chemmater.0c03057},
  DOI = {10.1021/acs.chemmater.0c03057},
  number = {18},
  journal = {Chem. Mater.},
  publisher = {American Chemical Society (ACS)},
  author = {Thyagarajan,  Raghuram and Sholl,  David S.},
  year = {2020},
  month = aug,
  pages = {8020–8033}
}

@article{SHAPIRO1965,
  title = {An analysis of variance test for normality (complete samples)},
  volume = {52},
  ISSN = {1464-3510},
  url = {http://dx.doi.org/10.1093/biomet/52.3-4.591},
  DOI = {10.1093/biomet/52.3-4.591},
  number = {3–4},
  journal = {Biometrika},
  publisher = {Oxford University Press (OUP)},
  author = {SHAPIRO,  S. S. and WILK,  M. B.},
  year = {1965},
  month = dec,
  pages = {591–611}
}

@article{Tanskanen2014,
  title = {Correlating Growth Characteristics in Atomic Layer Deposition with Precursor Molecular Structure: The Case of Zinc Tin Oxide},
  volume = {26},
  ISSN = {1520-5002},
  url = {http://dx.doi.org/10.1021/cm403913r},
  DOI = {10.1021/cm403913r},
  number = {9},
  journal = {Chem. Mater.},
  publisher = {American Chemical Society (ACS)},
  author = {Tanskanen,  Jukka T. and H\"{a}gglund,  Carl and Bent,  Stacey F.},
  year = {2014},
  month = apr,
  pages = {2795–2802}
}

@article{Lee2013,
  title = {Thermodynamic stability of various phases of zinc tin oxides from ab initio calculations},
  volume = {1},
  ISSN = {2050-7534},
  url = {http://dx.doi.org/10.1039/c3tc30960f},
  DOI = {10.1039/c3tc30960f},
  number = {39},
  journal = {J. Mater. Chem. C},
  publisher = {Royal Society of Chemistry (RSC)},
  author = {Lee,  Joohwi and Lee,  Seung-Cheol and Hwang,  Cheol Seong and Choi,  Jung-Hae},
  year = {2013},
  pages = {6364}
}

@article{Krner2015,
  title = {Generic origin of subgap states in transparent amorphous semiconductor oxides illustrated for the cases of In-Zn-O and In-Sn-O: Subgap states in transparent amorphous semiconductor oxides},
  volume = {212},
  ISSN = {1862-6300},
  url = {http://dx.doi.org/10.1002/pssa.201431871},
  DOI = {10.1002/pssa.201431871},
  number = {7},
  journal = {Phys. Status Solidi A},
  publisher = {Wiley},
  author = {K\"{o}rner,  Wolfgang and Urban,  Daniel F. and Els\"{a}sser,  Christian},
  year = {2015},
  month = mar,
  pages = {1476–1481}
}

@article{Liu2016,
  title = {First-principles calculations of structural,  elastic,  and electronic properties of trigonal ZnSnO3 under pressure},
  volume = {180},
  ISSN = {0254-0584},
  url = {http://dx.doi.org/10.1016/j.matchemphys.2016.05.041},
  DOI = {10.1016/j.matchemphys.2016.05.041},
  journal = {Mater. Chem. Phys.},
  publisher = {Elsevier BV},
  author = {Liu,  Qi-Jun and Qin,  Han and Jiao,  Zhen and Liu,  Fu-Sheng and Liu,  Zheng-Tang},
  year = {2016},
  month = sep,
  pages = {75–81}
}

@article{Husein2020,
  title = {The Role of Cation Coordination in the Electrical and Optical Properties of Amorphous Transparent Conducting Oxides},
  volume = {32},
  ISSN = {1520-5002},
  url = {http://dx.doi.org/10.1021/acs.chemmater.0c01672},
  DOI = {10.1021/acs.chemmater.0c01672},
  number = {15},
  journal = {Chem. Mater.},
  publisher = {American Chemical Society (ACS)},
  author = {Husein,  Sebastian and Medvedeva,  Julia E. and Perkins,  John D. and Bertoni,  Mariana I.},
  year = {2020},
  month = jun,
  pages = {6444–6455}
}

@inproceedings{Subhechha2022,
  doi = {10.1109/icicdt56182.2022.9933087},
  url = {https://doi.org/10.1109/icicdt56182.2022.9933087},
  year = {2022},
  month = sep,
  publisher = {{IEEE}},
  author = {S. Subhechha and N. Rassoul and A. Belmonte and H. Hody and H. Dekkers and M. J. van Setten and A. Chasin and S.H. Sharifi and K. Banerjee and H. Puliyalil and S. Kundu and M. Pak and D. Tsvetanova and N. Bazzazian and K. Vandersmissen and D. Batuk and J. Geypen and J. Heijlen and R. Delhougne and G. S. Kar},
  title = {Device engineering guidelines for performance boost in {IGZO} front gated {TFTs} based on defect control},
  booktitle = {2022 International Conference on {IC} Design and Technology ({ICICDT})}
}

@article{Kruv2023,
  doi = {10.1109/ted.2023.3297976},
  url = {https://doi.org/10.1109/ted.2023.3297976},
  year = {2023},
  month = sep,
  publisher = {Institute of Electrical and Electronics Engineers ({IEEE})},
  volume = {70},
  number = {9},
  pages = {4674--4679},
  author = {Anastasiia Kruv and M. J. Van Setten and Hendrik F. W. Dekkers and Christophe Lorant and Devin Verreck and Quentin Smets and Bhuvaneshwari Yengula Venkataramana and Attilio Belmonte and Subhali Subhechha and Adrian Vaisman Chasin and Romain Delhougne and Gouri Sankar Kar},
  title = {The Impact of {IGZO} Channel Composition on {DRAM} Transistor Performance},
  journal = {{IEEE} Trans. Electron Devices}
}

@article{Krner2013,
  doi = {10.1063/1.4826895},
  url = {https://doi.org/10.1063/1.4826895},
  year = {2013},
  month = oct,
  publisher = {{AIP} Publishing},
  volume = {114},
  number = {16},
  pages = {163704},
  author = {Wolfgang K\"{o}rner and Daniel F. Urban and Christian Els\"{a}sser},
  title = {Origin of subgap states in amorphous In-Ga-Zn-O},
  journal = {J. Appl. Phys.}
}

@INPROCEEDINGS{Subhechha21_oxygen_tunnel,
  author={Subhechha, S. and Rassoul, N. and Belmonte, A. and Delhougne, R. and Banerjee, K. and Donadio, G. L. and Dekkers, H. and van Setten, M. J. and Puliyalil, H. and Mao, M. and Kundu, S. and Pak, M. and Teugels, L. and Tsvetanova, D. and Bazzazian, N. and Klijs, L. and Hody, H. and Chasin, A. and Heijlen, J. and Goux, L. and Kar, G. S.},
  booktitle={2021 Symposium on VLSI Technology}, 
  title={First demonstration of sub-12 nm Lg gate last IGZO-TFTs with oxygen tunnel architecture for front gate devices}, 
  year={2021},
  volume={},
  number={},
  pages={1-2},
  doi={}}

@article{vanSetten2017,
  doi = {10.1103/physrevb.96.155207},
  url = {https://doi.org/10.1103/physrevb.96.155207},
  year = {2017},
  month = oct,
  publisher = {American Physical Society ({APS})},
  volume = {96},
  number = {15},
  pages = {155207},
  author = {M. J. van Setten and M. Giantomassi and X. Gonze and G.-M. Rignanese and G. Hautier},
  title = {Automation methodologies and large-scale validation for GW: Towards high-throughput GW calculations},
  journal = {Phys. Rev. B}
}

@article{Han2021,
  doi = {10.3390/electronics11010053},
  url = {https://doi.org/10.3390/electronics11010053},
  year = {2021},
  month = dec,
  publisher = {{MDPI} {AG}},
  volume = {11},
  number = {1},
  pages = {53},
  author = {Hoonhee Han and Seokmin Jang and Duho Kim and Taeheun Kim and Hyeoncheol Cho and Heedam Shin and Changhwan Choi},
  title = {Memory Characteristics of Thin Film Transistor with Catalytic Metal Layer Induced Crystallized Indium-Gallium-Zinc-Oxide ({IGZO}) Channel},
  journal = {Electron.}
}

@article{BROYDEN1970,
  doi = {10.1093/imamat/6.1.76},
  url = {https://doi.org/10.1093/imamat/6.1.76},
  year = {1970},
  publisher = {Oxford University Press ({OUP})},
  volume = {6},
  number = {1},
  pages = {76--90},
  author = {C. G. Broyden},
  title = {The Convergence of a Class of Double-rank Minimization Algorithms 1. General Considerations},
  journal = {IMA J. Appl. Math.}
}

@article{Fletcher1970,
  doi = {10.1093/comjnl/13.3.317},
  url = {https://doi.org/10.1093/comjnl/13.3.317},
  year = {1970},
  month = mar,
  publisher = {Oxford University Press ({OUP})},
  volume = {13},
  number = {3},
  pages = {317--322},
  author = {R. Fletcher},
  title = {A new approach to variable metric algorithms},
  journal = {Comput. J.}
}

@article{Goldfarb1970,
  doi = {10.1090/s0025-5718-1970-0258249-6},
  url = {https://doi.org/10.1090/s0025-5718-1970-0258249-6},
  year = {1970},
  publisher = {American Mathematical Society ({AMS})},
  volume = {24},
  number = {109},
  pages = {23--26},
  author = {Donald Goldfarb},
  title = {A family of variable-metric methods derived by variational means},
  journal = {Math. Comput.}
}

@article{Shanno1970,
  doi = {10.1090/s0025-5718-1970-0274029-x},
  url = {https://doi.org/10.1090/s0025-5718-1970-0274029-x},
  year = {1970},
  publisher = {American Mathematical Society ({AMS})},
  volume = {24},
  number = {111},
  pages = {647--656},
  author = {D. F. Shanno},
  title = {Conditioning of quasi-Newton methods for function minimization},
  journal = {Math. Comput.}
}

@article{Shiah2021,
  doi = {10.1038/s41928-021-00671-0},
  url = {https://doi.org/10.1038/s41928-021-00671-0},
  year = {2021},
  month = nov,
  publisher = {Springer Science and Business Media {LLC}},
  volume = {4},
  number = {11},
  pages = {800--807},
  author = {Yu-Shien Shiah and Kihyung Sim and Yuhao Shi and Katsumi Abe and Shigenori Ueda and Masato Sasase and Junghwan Kim and Hideo Hosono},
  title = {Mobility{\textendash}stability trade-off in oxide thin-film transistors},
  journal = {Nat. Electron.}
}

@article{Kamiya2010,
  doi = {10.1002/pssa.200983772},
  url = {https://doi.org/10.1002/pssa.200983772},
  year = {2010},
  month = may,
  publisher = {Wiley},
  volume = {207},
  number = {7},
  pages = {1698--1703},
  author = {Toshio Kamiya and Kenji Nomura and Hideo Hosono},
  title = {Subgap states,  doping and defect formation energies in amorphous oxide semiconductor a-{InGaZnO}4 studied by density functional theory},
  journal = {Phys. Status Solidi A}
}

@article{Park2020,
  doi = {10.1149/ma2020-02281930mtgabs},
  url = {https://doi.org/10.1149/ma2020-02281930mtgabs},
  year = {2020},
  month = nov,
  publisher = {The Electrochemical Society},
  volume = {{MA}2020-02},
  number = {28},
  pages = {1930--1930},
  author = {Ji-Min Park and Hyun-Suk Kim},
  title = {Ultra-High Mobility Transistors Via Metal Induced Conductive Region},
  journal = {{ECS} Meeting Abstracts}
}

@article{AvelarMuoz2020,
  doi = {10.1016/j.jallcom.2020.155353},
  url = {https://doi.org/10.1016/j.jallcom.2020.155353},
  year = {2020},
  month = sep,
  publisher = {Elsevier {BV}},
  volume = {835},
  pages = {155353},
  author = {F. Avelar-Mu{\~{n}}oz and J.A. Berumen and M.A. Aguilar-Frutis and J.J. Araiza and J.J. Ortega},
  title = {Enhancement on carrier mobility in amorphous indium tin oxynitride ({ITON}) thin films},
  journal = {J. Alloys Compd.}
}

@article{Jeong2020,
  doi = {10.3390/electronics9111875},
  url = {https://doi.org/10.3390/electronics9111875},
  year = {2020},
  month = nov,
  publisher = {{MDPI} {AG}},
  volume = {9},
  number = {11},
  pages = {1875},
  author = {Hwan-Seok Jeong and Hyun Seok Cha and Seong Hyun Hwang and Hyuck-In Kwon},
  title = {Effects of Annealing Atmosphere on Electrical Performance and Stability of High-Mobility Indium-Gallium-Tin Oxide Thin-Film Transistors},
  journal = {Electron.}
}

@article{Guo2020,
  doi = {10.1063/1.5140234},
  url = {https://doi.org/10.1063/1.5140234},
  year = {2020},
  month = sep,
  publisher = {{AIP} Publishing},
  volume = {10},
  number = {9},
  pages = {095317},
  author = {Hong-Bo Guo and Fei Shan and Han-Sang Kim and Jae-Yun Lee and Nam Kim and Yu Zhao and Sung-Jin Kim},
  title = {Amorphous oxide thin-film transistors and inverters enabled by solution-processed multi-layers as active channels},
  journal = {AIP Adv.}
}

@article{Lestari2020,
  doi = {10.1166/jnn.2020.17222},
  url = {https://doi.org/10.1166/jnn.2020.17222},
  year = {2020},
  month = jan,
  publisher = {American Scientific Publishers},
  volume = {20},
  number = {1},
  pages = {252--256},
  author = {Annisa Dwi Lestari and Maryane Putri and Young-Woo Heo and Hee Young Lee},
  title = {Influence of Oxygen Partial Pressure on Radio Frequency Magnetron Sputtered Amorphous {InZnSnO} Thin Film Transistors},
  journal = {J. Nanosci. Nanotechnol.}
}

@article{Cha2020,
  doi = {10.3390/electronics9122196},
  url = {https://doi.org/10.3390/electronics9122196},
  year = {2020},
  month = dec,
  publisher = {{MDPI} {AG}},
  volume = {9},
  number = {12},
  pages = {2196},
  author = {Hyun-Seok Cha and Hwan-Seok Jeong and Seong-Hyun Hwang and Dong-Ho Lee and Hyuck-In Kwon},
  title = {Electrical Performance and Stability Improvements of High-Mobility Indium{\textendash}Gallium{\textendash}Tin Oxide Thin-Film Transistors Using an Oxidized Aluminum Capping Layer of Optimal Thickness},
  journal = {Electron.}
}

@article{Takahashi2020,
  doi = {10.35848/1882-0786/ab88c5},
  url = {https://doi.org/10.35848/1882-0786/ab88c5},
  year = {2020},
  month = apr,
  publisher = {{IOP} Publishing},
  volume = {13},
  number = {5},
  pages = {054003},
  author = {Takanori Takahashi and Mami N. Fujii and Ryoko Miyanaga and Miki Miyanaga and Yasuaki Ishikawa and Yukiharu Uraoka},
  title = {Unique degradation under {AC} stress in high-mobility amorphous In{\textendash}W{\textendash}Zn{\textendash}O thin-film transistors},
  journal = {Appl. Phys. Express}
}

@article{Choi2020,
  doi = {10.1109/ted.2020.2968592},
  url = {https://doi.org/10.1109/ted.2020.2968592},
  year = {2020},
  month = mar,
  publisher = {Institute of Electrical and Electronics Engineers ({IEEE})},
  volume = {67},
  number = {3},
  pages = {1014--1020},
  author = {Il Man Choi and Min Jae Kim and Nuri On and Aeran Song and Kwun-Bum Chung and Hoon Jeong and Jeong Ki Park and Jae Kyeong Jeong},
  title = {Achieving High Mobility and Excellent Stability in Amorphous In{\textendash}Ga{\textendash}Zn{\textendash}Sn{\textendash}O Thin-Film Transistors},
  journal = {{IEEE} Trans. Electron Devices}
}

@article{Moffitt2017,
  doi = {10.1002/aelm.201700189},
  url = {https://doi.org/10.1002/aelm.201700189},
  year = {2017},
  month = sep,
  publisher = {Wiley},
  volume = {3},
  number = {10},
  pages = {1700189},
  author = {Stephanie L. Moffitt and Qimin Zhu and Qing Ma and Allison F. Falduto and D. Bruce Buchholz and Robert P. H. Chang and Thomas O. Mason and Julia E. Medvedeva and Tobin J. Marks and Michael J. Bedzyk},
  title = {Probing the Unique Role of Gallium in Amorphous Oxide Semiconductors through Structure{\textendash}Property Relationships},
  journal = {Adv. Electron. Mater.}
}

@article{Medvedeva2017,
  doi = {10.1002/aelm.201700082},
  url = {https://doi.org/10.1002/aelm.201700082},
  year = {2017},
  month = aug,
  publisher = {Wiley},
  volume = {3},
  number = {9},
  pages = {1700082},
  author = {Julia E. Medvedeva and D. Bruce Buchholz and Robert P. H. Chang},
  title = {Recent Advances in Understanding the Structure and Properties of Amorphous Oxide Semiconductors},
  journal = {Adv. Electron. Mater.}
}

@article{Zhang2020,
  doi = {10.1063/5.0032897},
  url = {https://doi.org/10.1063/5.0032897},
  year = {2020},
  month = dec,
  publisher = {{AIP} Publishing},
  volume = {128},
  number = {21},
  pages = {215704},
  author = {Zhaofu Zhang and Yuzheng Guo and John Robertson},
  title = {Role of the third metal oxide in In{\textendash}Ga{\textendash}Zn{\textendash}O4 amorphous oxide semiconductors: Alternatives to gallium},
  journal = {J. Appl. Phys.}
}

@article{Sheng1980,
  title = {Fluctuation-induced tunneling conduction in disordered materials},
  volume = {21},
  ISSN = {0163-1829},
  url = {http://dx.doi.org/10.1103/PhysRevB.21.2180},
  DOI = {10.1103/physrevb.21.2180},
  number = {6},
  journal = {Phys. Rev. B},
  publisher = {American Physical Society (APS)},
  author = {Sheng,  Ping},
  year = {1980},
  month = mar,
  pages = {2180–2195}
}

@article{Lordi2010,
  title = {Charge carrier scattering by defects in semiconductors},
  volume = {81},
  ISSN = {1550-235X},
  url = {http://dx.doi.org/10.1103/PhysRevB.81.235204},
  DOI = {10.1103/physrevb.81.235204},
  number = {23},
  pages = {235204},
  journal = {Phys. Rev. B},
  publisher = {American Physical Society (APS)},
  author = {Lordi,  Vincenzo and Erhart,  Paul and Åberg,  Daniel},
  year = {2010},
  month = jun 
}

@article{Guo2024,
  title = {Total-Ionizing-Dose Effects in IGZO Thin-Film Transistors With SiO$_2$ Oxygen-Penetration Layers},
  volume = {71},
  ISSN = {1558-1578},
  url = {http://dx.doi.org/10.1109/TNS.2023.3346825},
  DOI = {10.1109/tns.2023.3346825},
  number = {4},
  journal = {IEEE Transactions on Nuclear Science},
  publisher = {Institute of Electrical and Electronics Engineers (IEEE)},
  author = {Guo,  Zixiang and Zhang,  En Xia and Chasin,  A. and Linten,  D. and Belmonte,  A. and Kar,  G. and Reed,  Robert A. and Schrimpf,  Ronald D. and Fleetwood,  Daniel M.},
  year = {2024},
  month = apr,
  pages = {461–468}
}

@article{Kong2024,
  title = {Discovering the Impact of Cooling Scheme During Annealing: A New Knob for Achieving Thermally Stable IGZO FETs},
  volume = {71},
  ISSN = {1557-9646},
  url = {http://dx.doi.org/10.1109/TED.2024.3433832},
  DOI = {10.1109/ted.2024.3433832},
  number = {9},
  journal = {IEEE Trans. Electron Devices},
  publisher = {Institute of Electrical and Electronics Engineers (IEEE)},
  author = {Kong,  Qiwen and Liu,  Long and Han,  Kaizhen and Sun,  Chen and Jiao,  Leming and Zhou,  Zuopu and Zheng,  Zijie and Liu,  Gan and Xu,  Haiwen and Zhang,  Jishen and Chen,  Yue and Gong,  Xiao},
  year = {2024},
  month = sep,
  pages = {5425–5431}
}

@article{Janotti2009,
  doi = {10.1088/0034-4885/72/12/126501},
  url = {https://doi.org/10.1088/0034-4885/72/12/126501},
  year = {2009},
  month = oct,
  publisher = {{IOP} Publishing},
  volume = {72},
  number = {12},
  pages = {126501},
  author = {Anderson Janotti and Chris G Van de Walle},
  title = {Fundamentals of zinc oxide as a semiconductor},
  journal = {Rep. Prog. Phys.}
}

@inproceedings{Kljucar2020,
  series = {SSDM2020},
  title = {300mm IGZO nFETs with low-T Ru contacts for localized doping and increased BEOL compatibility},
  url = {http://dx.doi.org/10.7567/SSDM.2020.J-6-03},
  DOI = {10.7567/ssdm.2020.j-6-03},
  booktitle = {Extended Abstracts of the 2020 International Conference on Solid State Devices and Materials},
  publisher = {The Japan Society of Applied Physics},
  author = {Kljucar,  Luka and Smets,  Quentin and {van Setten},  Michiel and Mitard,  Jerome and Belmonte,  Atillio and Dekkers,  Harold and Teugels,  Lieve and Mao,  Ming and Puliyalil,  Harinarayanan and Borniquel,  Jose Ignaciodel Agua and Delhougne,  Romain and Kar,  Gouri Sankar and Tokei,  Zsolt},
  year = {2020},
  month = sep,
  collection = {SSDM2020}
}

@article{vanSetten2021_odef_igzo,
  doi = {10.1021/acsaelm.1c00553},
  url = {https://doi.org/10.1021/acsaelm.1c00553},
  year = {2021},
  pages = {4037–4046},
  month = sep,
  publisher = {American Chemical Society ({ACS})},
  author = {Michiel J. {van Setten} and Harold F. W. Dekkers and Luka Kljucar and Jerome Mitard and Christopher Pashartis and Subhali Subhechha and Nouredine Rassoul and Romain Delhougne and Gouri S. Kar and Geoffrey Pourtois},
  title = {Oxygen Defect Stability in Amorphous,  C-Axis Aligned,  and Spinel {IGZO}},
  journal = {ACS Appl. Electron. Mater.}
}

@article{vanSetten2024_igzo_h,
  doi = {},
  url = {},
  year = {2024},
  pages = {},
  publisher = {},
  author = {Michiel J. {van Setten} and Harold F. W. Dekkers and Geoffrey Pourtois},
  title = {Hydrogen binding in Amorphous,  C-Axis Aligned,  and Spinel {IGZO}},
  journal = {in preparation}
}

@article{vanSetten2022screening,
  doi = {10.1039/d2ma00759b},
  url = {https://doi.org/10.1039/d2ma00759b},
  year = {2022},
  publisher = {Royal Society of Chemistry ({RSC})},
  volume = {3},
  number = {23},
  pages = {8413--8427},
  author = {Michiel J. {van Setten} and Hendrik F. W. Dekkers and Christopher Pashartis and Adrian Chasin and Attilio Belmonte and Romain Delhougne and Gouri S. Kar and Geoffrey Pourtois},
  title = {Complex amorphous oxides: property prediction from high throughput {DFT} and {AI} for new material search},
  journal = {Materials Advances}
}

@inproceedings{Belmonte2020,
  doi = {10.1109/iedm13553.2020.9371900},
  url = {https://doi.org/10.1109/iedm13553.2020.9371900},
  year = {2020},
  month = dec,
  publisher = {{IEEE}},
  author = {A. Belmonte and H. Oh and N. Rassoul and G.L. Donadio and J. Mitard and H. Dekkers and R. Delhougne and S. Subhechha and A. Chasin and M. J. {van Setten} and L. Kljucar and M. Mao and H. Puliyalil and M. Pak and L. Teugels and D. Tsvetanova and K. Banerjee and L. Souriau and Z. Tokei and L. Goux and G. S. Kar},
  title = {Capacitor-less,  Long-Retention ($>$400s) {DRAM} Cell Paving the Way towards Low-Power and High-Density Monolithic 3D {DRAM}},
  booktitle = {2020 {IEEE} International Electron Devices Meeting ({IEDM})}
}

@article{Kang2021,
  doi = {10.1016/j.materresbull.2021.111252},
  url = {https://doi.org/10.1016/j.materresbull.2021.111252},
  year = {2021},
  month = jul,
  publisher = {Elsevier {BV}},
  volume = {139},
  pages = {111252},
  author = {Youngjin Kang and Woobin Lee and Jaeyoung Kim and Kyobin Keum and Seung-Han Kang and Jeong-Wan Jo and Sung Kyu Park and Yong-Hoon Kim},
  title = {Effects of crystalline structure of {IGZO} thin films on the electrical and photo-stability of metal-oxide thin-film transistors},
  journal = {Mater. Res. Bull.}
}

@article{Vogt2020,
  doi = {10.1103/physrevresearch.2.033358},
  url = {https://doi.org/10.1103/physrevresearch.2.033358},
  year = {2020},
  month = sep,
  publisher = {American Physical Society ({APS})},
  volume = {2},
  number = {3},
  pages = {033358},
  author = {Kyle T. Vogt and Christopher E. Malmberg and Jacob C. Buchanan and George W. Mattson and G. Mirek Brandt and Dylan B. Fast and Paul H.-Y. Cheong and John F. Wager and Matt W. Graham},
  title = {Ultrabroadband density of states of amorphous In-Ga-Zn-O},
  journal = {Phys. Rev. Research}
}

@article{Kim2014,
  doi = {10.7567/jjap.53.08ng03},
  url = {https://doi.org/10.7567/jjap.53.08ng03},
  year = {2014},
  month = jul,
  publisher = {{IOP} Publishing},
  volume = {53},
  number = {8S3},
  pages = {08NG03},
  author = {Ju-Yeon Kim and So Hyeon Jeong and Kyeong Min Yu and Eui-Jung Yun and Byung Seong Bae},
  title = {Stabilities of amorphous indium gallium zinc oxide thin films under light illumination with various wavelengths and intensities},
  journal = {Japanese J. Appl. Phys.}
}

@inproceedings{Hiblot2021,
  doi = {10.1109/irps46558.2021.9405201},
  url = {https://doi.org/10.1109/irps46558.2021.9405201},
  year = {2021},
  month = mar,
  publisher = {{IEEE}},
  author = {Gaspard Hiblot and Nouredine Rassoul and Lieve Teugels and Katia Devriendt and Adrian Vaisman Chasin and Michiel J. {van Setten} and Attilio Belmonte and Romain Delhougne and Gouri Sankar Kar},
  title = {Process-induced charging damage in {IGZO} {nTFTs}},
  booktitle = {2021 {IEEE} International Reliability Physics Symposium ({IRPS})}
}

@article{Mitard2020,
  doi = {10.1149/09807.0205ecst},
  url = {https://doi.org/10.1149/09807.0205ecst},
  year = {2020},
  month = sep,
  publisher = {The Electrochemical Society},
  volume = {98},
  number = {7},
  pages = {205--217},
  author = {Jerome Mitard and Luka Kljucar and Nouredine Rassoul and Harold F. W. Dekkers and Michiel {van Setten} and Adrian Chasin and Geoffrey Pourtois and Attilio Belmonte and Romain Delhougne and Gabriele Luca Donadio and Ludovic Goux and Manoj Nag and Chris Wilson and Zsolt Tokei and Jose Ignacio del agua Borniquel and Soeren Steudel and Gouri Sankar Kar},
  title = {(Invited) Sub-40mV Sigma {VTH} Igzo {nFETs} in 300mm Fab},
  journal = {{ECS} Transactions}
}

@article{DeRoose2017,
  doi = {10.1109/jssc.2017.2731808},
  url = {https://doi.org/10.1109/jssc.2017.2731808},
  year = {2017},
  month = nov,
  publisher = {Institute of Electrical and Electronics Engineers ({IEEE})},
  volume = {52},
  number = {11},
  pages = {3095--3103},
  author = {Florian De Roose and Kris Myny and Marc Ameys and Jan-Laurens P. J. van der Steen and Joris Maas and Joris de Riet and Jan Genoe and Wim Dehaene},
  title = {A Thin-Film,  a-{IGZO},  128b {SRAM} and {LPROM} Matrix With Integrated Periphery on Flexible Foil},
  journal = { IEEE J. Solid-State Circuits}
}

@article{Chasin2014,
  doi = {10.1109/led.2014.2314704},
  url = {https://doi.org/10.1109/led.2014.2314704},
  year = {2014},
  month = jun,
  publisher = {Institute of Electrical and Electronics Engineers ({IEEE})},
  volume = {35},
  number = {6},
  pages = {642--644},
  author = {Adrian Chasin and Leqi Zhang and Ajay Bhoolokam and Manoj Nag and Soeren Steudel and Bogdan Govoreanu and Georges Gielen and Paul Heremans},
  title = {High-Performance a-{IGZO} Thin Film Diode as Selector for Cross-Point Memory Application},
  journal = {IEEE Electron Device Lett.}
}

@article{deJamblinnedeMeux2018-iswo,
  doi = {10.1103/physrevb.97.045208},
  url = {https://doi.org/10.1103/physrevb.97.045208},
  year = {2018},
  month = {jan},
  publisher = {American Physical Society ({APS})},
  volume = {97},
  pages = {045208},
  number = {4},
  author = {A. de Jamblinne de Meux and G. Pourtois and J. Genoe and P. Heremans},
  title = {Method to quantify the delocalization of electronic states in amorphous semiconductors and its application to assessing charge carrier mobility of P-type amorphous oxide semiconductors},
  journal = {Phys. Rev. B}
}

@article{Bortnik2014,
 author = {Borštnik, Urban and VandeVondele, Joost and Weber, Valéry and Hutter, Jürg},
 doi = {10.1016/j.parco.2014.03.012},
 url = {https://doi.org/10.1016/j.parco.2014.03.012},
 year = {2014},
 month = may,
 publisher = {Elsevier BV},
 volume = {40},
 number = {5-6},
 pages = {47-58},
 title = {Sparse matrix multiplication: {The} distributed block-compressed sparse row library},
 journal = {Parallel Comput.},
 source = {Crossref},
 issn = {0167-8191},
}

@article{Khne2020,
 author = {Kühne, Thomas D. and Iannuzzi, Marcella and Del Ben, Mauro and Rybkin, Vladimir V. and Seewald, Patrick and Stein, Frederick and Laino, Teodoro and Khaliullin, Rustam Z. and Schütt, Ole and Schiffmann, Florian and Golze, Dorothea and Wilhelm, Jan and Chulkov, Sergey and Bani-Hashemian, Mohammad Hossein and Weber, Valéry and Borštnik, Urban and Taillefumier, Mathieu and Jakobovits, Alice Shoshana and Lazzaro, Alfio and Pabst, Hans and Müller, Tiziano and Schade, Robert and Guidon, Manuel and Andermatt, Samuel and Holmberg, Nico and Schenter, Gregory K. and Hehn, Anna and Bussy, Augustin and Belleflamme, Fabian and Tabacchi, Gloria and Glöß, Andreas and Lass, Michael and Bethune, Iain and Mundy, Christopher J. and Plessl, Christian and Watkins, Matt and VandeVondele, Joost and Krack, Matthias and Hutter, Jürg},
 doi = {10.1063/5.0007045},
 url = {https://doi.org/10.1063/5.0007045},
 year = {2020},
 month = may,
 publisher = {AIP Publishing},
 volume = {152},
 number = {19},
 pages = {194103},
 title = {{CP2K:} {An} electronic structure and molecular dynamics software package - Quickstep: {Efficient} and accurate electronic structure calculations},
 journal = {J. Chem. Phys.},
 source = {Crossref},
 issn = {0021-9606, 1089-7690},
}

@article{Hutter2013,
 author = {Hutter, Jürg and Iannuzzi, Marcella and Schiffmann, Florian and VandeVondele, Joost},
 doi = {10.1002/wcms.1159},
 url = {https://doi.org/10.1002/wcms.1159},
 year = {2014},
 month = jun,
 publisher = {Wiley},
 volume = {4},
 number = {1},
 pages = {15-25},
 title = {cp2k: {Atomistic} simulations of condensed matter systems},
 journal = {WIREs Comput Mol Sci},
 source = {Crossref},
 subtitle = {cp
2k
 Simulation Software},
 issn = {1759-0876},
}

@article{Perdew2008,
 author = {Perdew, John P. and Ruzsinszky, Adrienn and Csonka, Gábor I. and Vydrov, Oleg A. and Scuseria, Gustavo E. and Constantin, Lucian A. and Zhou, Xiaolan and Burke, Kieron},
 doi = {10.1103/physrevlett.100.136406},
 url = {https://doi.org/10.1103/physrevlett.100.136406},
 year = {2008},
 month = apr,
 publisher = {American Physical Society (APS)},
 volume = {100},
 number = {13},
 pages = {136406},
 title = {Restoring the Density-Gradient Expansion for Exchange in Solids and Surfaces},
 journal = {Phys. Rev. Lett.},
 source = {Crossref},
 issn = {0031-9007, 1079-7114},
}

@article{VandeVondele2007,
 author = {VandeVondele, Joost and Hutter, Jürg},
 doi = {10.1063/1.2770708},
 url = {https://doi.org/10.1063/1.2770708},
 year = {2007},
 month = sep,
 publisher = {AIP Publishing},
 volume = {127},
 number = {11},
 pages = {114105},
 title = {{Gaussian} basis sets for accurate calculations on molecular systems in gas and condensed phases},
 journal = {J. Chem. Phys.},
 source = {Crossref},
 issn = {0021-9606, 1089-7690},
}

@article{Krack2005,
 author = {Krack, M.},
 doi = {10.1007/s00214-005-0655-y},
 url = {https://doi.org/10.1007/s00214-005-0655-y},
 year = {2005},
 month = may,
 publisher = {Springer Science and Business Media LLC},
 volume = {114},
 number = {1-3},
 pages = {145-152},
 title = {Pseudopotentials for H to {Kr} optimized for gradient-corrected exchange-correlation functionals},
 journal = {Theor Chem Acc},
 source = {Crossref},
 issn = {1432-881X, 1432-2234},
}

@article{VandeVondele2005,
 author = {VandeVondele, Joost and Krack, Matthias and Mohamed, Fawzi and Parrinello, Michele and Chassaing, Thomas and Hutter, Jürg},
 doi = {10.1016/j.cpc.2004.12.014},
 url = {https://doi.org/10.1016/j.cpc.2004.12.014},
 year = {2005},
 month = apr,
 publisher = {Elsevier BV},
 volume = {167},
 number = {2},
 pages = {103-128},
 title = {Quickstep: {Fast} and accurate density functional calculations using a mixed {Gaussian} and plane waves approach},
 journal = {Comput. Phys. Commun.},
 source = {Crossref},
 issn = {0010-4655},
}

@article{Frigo2005,
 author = {Frigo, M. and Johnson, S.G.},
 doi = {10.1109/jproc.2004.840301},
 url = {https://doi.org/10.1109/jproc.2004.840301},
 year = {2005},
 month = feb,
 publisher = {Institute of Electrical and Electronics Engineers (IEEE)},
 volume = {93},
 number = {2},
 pages = {216-231},
 title = {The Design and Implementation of {FFTW3}},
 journal = {Proc. IEEE},
 source = {Crossref},
 issn = {0018-9219},
}

@article{Hartwigsen1998,
 author = {Hartwigsen, C. and Goedecker, S. and Hutter, J.},
 doi = {10.1103/physrevb.58.3641},
 url = {https://doi.org/10.1103/physrevb.58.3641},
 year = {1998},
 month = aug,
 publisher = {American Physical Society (APS)},
 volume = {58},
 number = {7},
 pages = {3641-3662},
 title = {Relativistic separable dual-space {Gaussian} pseudopotentials from H to {Rn}},
 journal = {Phys. Rev. B},
 source = {Crossref},
 issn = {0163-1829, 1095-3795},
}

@article{LIPPERT1997,
 author = {Lippert, Gerald and Hutter, Jurg and Parrinello, Michele},
 doi = {10.1080/00268979709482119},
 url = {https://doi.org/10.1080/00268979709482119},
 year = {1997},
 month = oct,
 publisher = {Informa UK Limited},
 volume = {92},
 number = {3},
 pages = {477-487},
 title = {A hybrid {Gaussian} and plane wave density functional scheme},
 journal = {Mol. Phys.},
 source = {Crossref},
 issn = {0026-8976},
}

@article{Perdew1996,
 author = {Perdew, John P. and Burke, Kieron and Ernzerhof, Matthias},
 doi = {10.1103/physrevlett.77.3865},
 url = {https://doi.org/10.1103/physrevlett.77.3865},
 year = {1996},
 month = oct,
 publisher = {American Physical Society (APS)},
 volume = {77},
 number = {18},
 pages = {3865-3868},
 title = {Generalized Gradient Approximation Made Simple},
 journal = {Phys. Rev. Lett.},
 source = {Crossref},
 issn = {0031-9007, 1079-7114},
}

\section{Supplementary material}
\setcounter{figure}{0}    
\renewcommand{\thefigure}{S\arabic{figure}}

\subsection{Data and data processing}

The full data and the jupyter notebook used to perform the data processing and analysis are provided separately as json and ipynb files.

\subsection{Radial distribution functions amorphous models}

\begin{figure}[p]
\centering
\includegraphics[width=0.67\columnwidth]{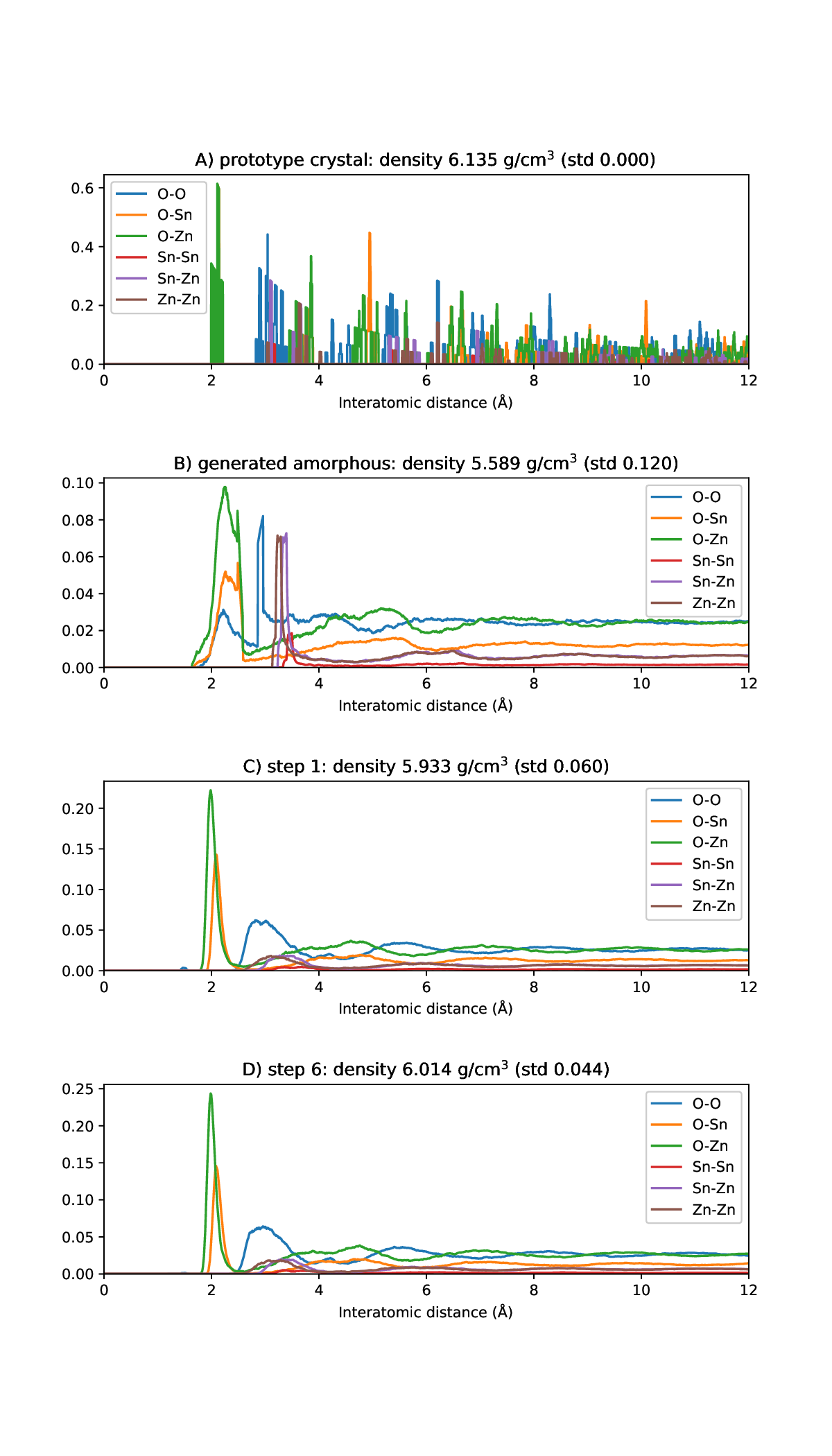}
\caption{Radial distribution function of the amorphous models of \ch{Zn2SnO4}, averaged over all models of this composition. The panels follow the steps of the structure generation and optimization. From top to bottom they contain the, A: prototype crystal structure, B: the initially generated random decorated models, C: the first BFGS optimized structures, D: the final optimized models including 2 cycles of TFMC. The final optimized models follow all features of the crystalline prototype except for a very small numbers of oxygen atoms getting closer to each other.}
\label{fig:s_rdf_21}
\end{figure}

\begin{figure}[p]
\centering
\includegraphics[width=0.67\columnwidth]{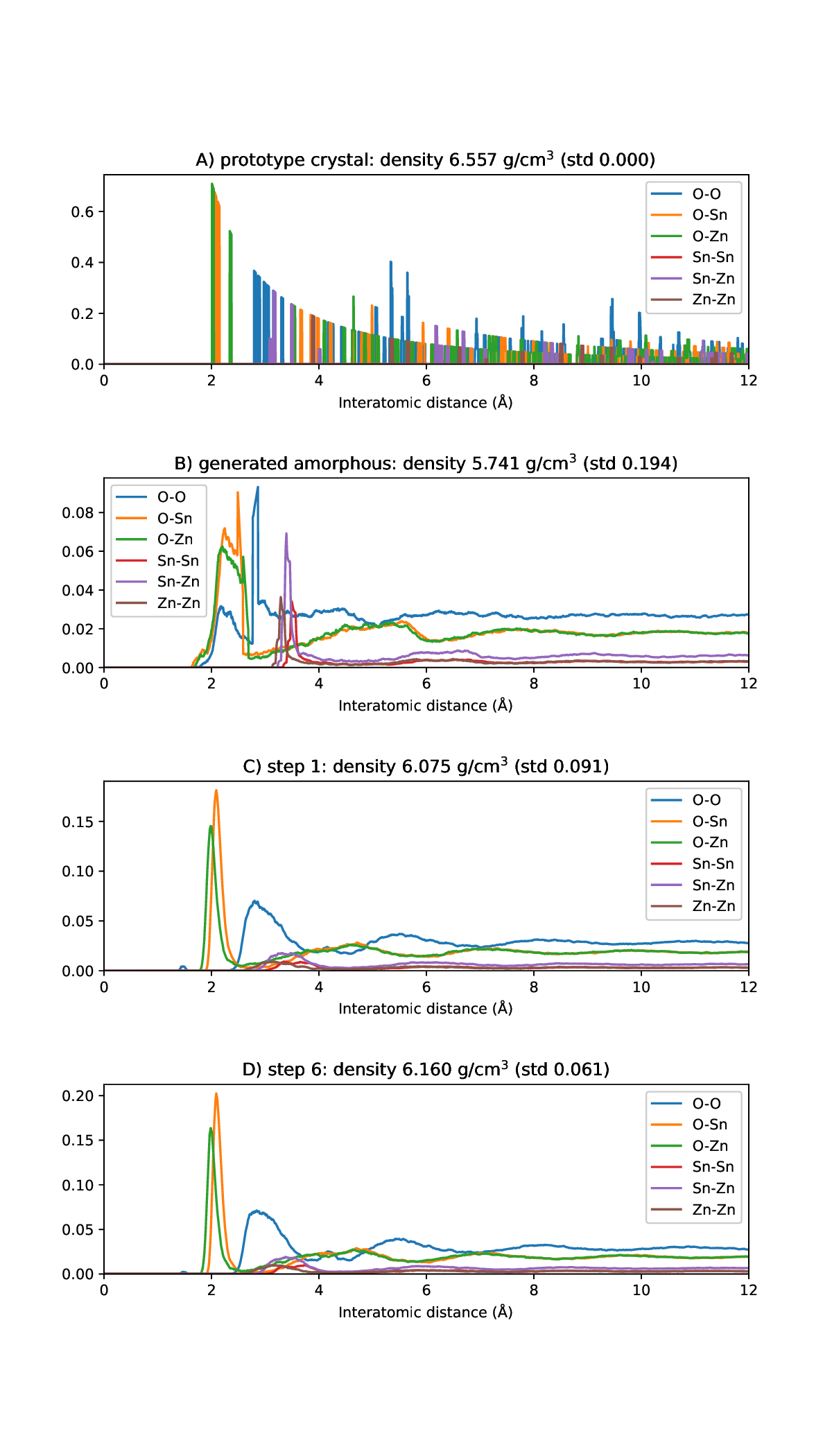}
\caption{Radial distribution function of the amorphous models of \ch{ZnSnO3}, averaged over all models of this composition. The panels follow the steps of the structure generation and optimization. From top to bottom they contain the, A: prototype crystal structure, B: the initially generated random decorated models, C: the first BFGS optimized structures, D: the final optimized models including 2 cycles of TFMC. The final optimized models follow all features of the crystalline prototype except for a very small numbers of oxygen atoms getting closer to each other.}
\label{fig:s_rdf_11}
\end{figure}

\begin{figure}[p]
\centering
\includegraphics[width=0.67\columnwidth]{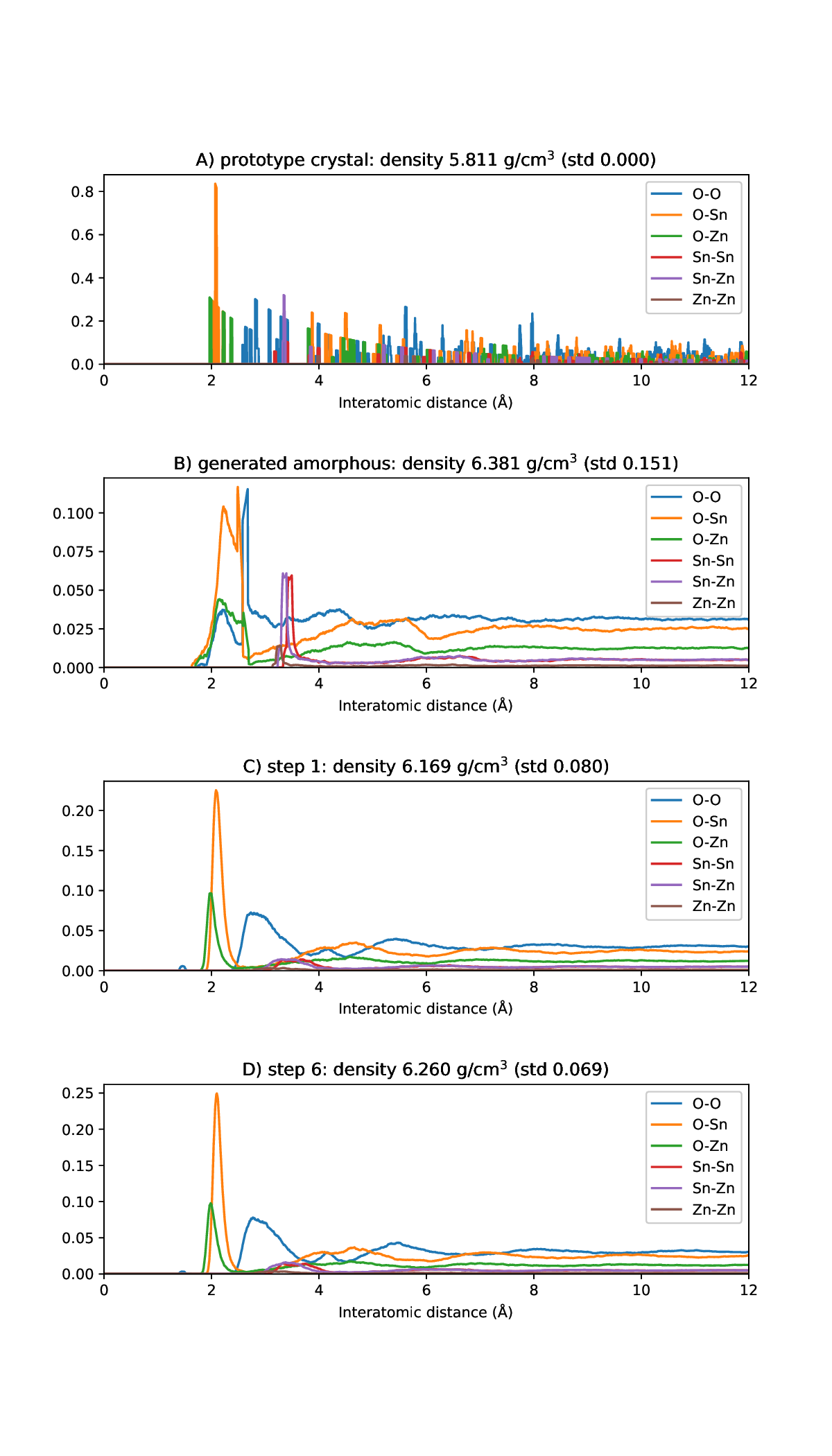}
\caption{Radial distribution function of the amorphous models of \ch{ZnSn2O5}, averaged over all models of this composition. The panels follow the steps of the structure generation and optimization. From top to bottom they contain the, A: prototype crystal structure, B: the initially generated random decorated models, C: the first BFGS optimized structures, D: the final optimized models including 2 cycles of TFMC. The final optimized models follow all features of the crystalline prototype except for a very small numbers of oxygen atoms getting closer to each other.}
\label{fig:s_rdf_12}
\end{figure}

\begin{figure}[p]
\centering
\includegraphics[width=0.67\columnwidth]{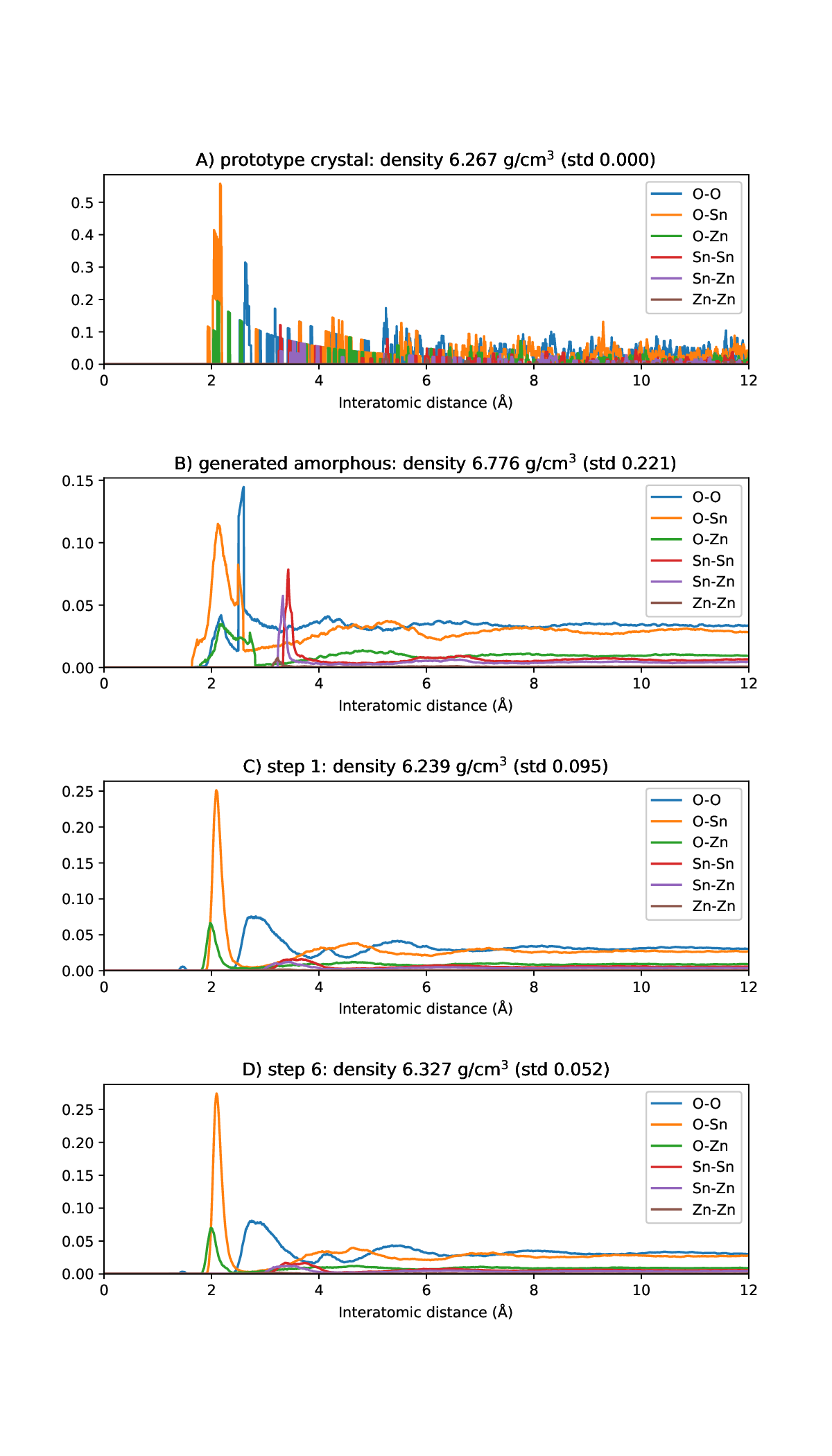}
\caption{Radial distribution function of the amorphous models of \ch{ZnSn3O7}, averaged over all models of this composition. The panels follow the steps of the structure generation and optimization. From top to bottom they contain the, A: prototype crystal structure, B: the initially generated random decorated models, C: the first BFGS optimized structures, D: the final optimized models including 2 cycles of TFMC. The final optimized models follow all features of the crystalline prototype except for a very small numbers of oxygen atoms getting closer to each other.}
\label{fig:s_rdf_13}
\end{figure}

\end{document}